\documentclass[preprint]{aastex}
\usepackage{epsf}

\def\linebreak{\hfil\break}

\def\
{\hfil\linebreak}

\def\degree{\ifmmode {^\circ}\else {$^\circ$}\fi}
\def\mum{\ifmmode {\rm \mu {\rm m}}\else $\rm \mu {\rm m}$\fi}
\def\arcsec{\ifmmode ^{\prime \prime}\else $^{\prime \prime}$\fi}

\def\inch{\ifmmode ^{\prime \prime}\else $^{\prime \prime}$\fi}
\def\arcmin{\ifmmode ^{\prime}\else $^{\prime}$\fi}

\def\msun{\ifmmode {\rm M_{\odot}}\else $\rm M_{\odot}$\fi}
\newbox\grsign \setbox\grsign=\hbox{$>$} \newdimen\grdimen \grdimen=\ht\grsign
\newbox\simlessbox \newbox\simgreatbox
\setbox\simgreatbox=\hbox{\raise.5ex\hbox{$>$}\llap
     {\lower.5ex\hbox{$\sim$}}}\ht1=\grdimen\dp1=0pt
\setbox\simlessbox=\hbox{\raise.5ex\hbox{$<$}\llap
     {\lower.5ex\hbox{$\sim$}}}\ht2=\grdimen\dp2=0pt

\def\simless{\mathrel{\copy\simlessbox}}

\begin{document}


\title{Planet Formation in the Outer Solar System}
\vskip 7ex
\author{Scott J. Kenyon}
\affil{Smithsonian Astrophysical Observatory,
60 Garden Street, Cambridge, MA 02138} 
\email{e-mail: skenyon@cfa.harvard.edu}


%
%

\begin{abstract}

This paper reviews coagulation models for planet formation in the 
Kuiper Belt, emphasizing links to recent observations of our and 
other solar systems.  At heliocentric distances of 35--50 AU, 
single annulus and multiannulus planetesimal accretion calculations 
produce several 1000 km or larger planets and many 50--500 km 
objects on timescales of 10--30 Myr in a Minimum Mass Solar Nebula.  
Planets form more rapidly in more massive nebulae.  All models 
yield two power law cumulative size distributions, 
$N_C \propto r^{-q}$ with $q$ = 3.0--3.5 for radii $r \gtrsim$ 10 km 
and $N_C \propto r^{-2.5}$ for radii $r \lesssim$ 1 km.  These size 
distributions are consistent with observations of Kuiper Belt objects 
acquired during the past decade.  Once large objects form at 35--50 AU, 
gravitational stirring leads to a collisional cascade where 0.1--10 km 
objects are ground to dust.  The collisional cascade removes 80\% 
to 90\% of the initial mass in the nebula in $\sim$ 1 Gyr.  This dust 
production rate is comparable to rates inferred for $\alpha$ Lyr, 
$\beta$ Pic, and other extrasolar debris disk systems.

\end{abstract}

\subjectheadings{planetary systems -- solar system: formation -- 
stars: formation -- circumstellar matter}

\section{INTRODUCTION}

Recent observations indicate that nearly all low and intermediate mass
stars are born with massive
circumstellar disks of gas and dust.  Most young pre-main sequence stars
with ages of $\sim$ 1 Myr have gaseous disks with sizes of 100 AU or 
larger and masses of $\sim$ 0.01 $M_{\odot}$ 
\citep{bec99,lad99}.  
Many older main sequence stars have dusty debris disks with sizes of 
100--1000 AU \citep{aum84,smi84,gai99,hab99,so00,spa01}.  
Current source statistics suggest the percentage of stars with observable 
disks declines from $\sim$ 100\% among the youngest stars to less than
10\% for stars more than 1 Gyr old \citep{bac93,art97,lad99,lag00}.

Models for the formation of our solar system naturally begin with a disk.
In the 1700's, Immanuel Kant and the Marquis de Laplace proposed that the 
solar system collapsed from a gaseous medium of roughly uniform density 
\citep{kan55,lap96}.  A flattened gaseous disk -- the protosolar nebula -- 
formed out of this cloud.  The Sun contracted out of material at the 
center of the disk; the planets condensed in the outer portions.  
Although other ideas have been studied since Laplace's time, this picture 
has gained widespread acceptance.  Measurements of the composition of the
earth, moon, and meteorites support a common origin for the sun and planets
\citep[e.g.,][]{har76,and89}.  Simulations of planet formation in a disk 
produce objects resembling known planets on timescales similar to the 
estimated lifetime of the protosolar nebula 
\citep{saf69,gre78,wet93,pol96,ale98,lev98,kok00,kor00,cha01}.

The Kuiper Belt provides a stern test of planet formation models.
In the past decade, observations have revealed several hundred objects 
with radii of 50--500 km in the ecliptic plane at distances of $\sim$ 
35--50 AU from the Sun \citep{jew93,luu96,gla97,jew98,chi99,luu00,gla01}.  
The total mass in these KBOs, $\sim$ 0.1 $M_{\oplus}$, suggests a 
reservoir of material left over from the formation of our solar system 
\citep{edg49,kui51}.  However, this mass is insufficient to allow the
formation of 500 km or larger KBOs on timescales of $\sim$ 5 Gyr
\citep{fer81,ste95,st97a,kl98}.

The Kuiper Belt also provides an interesting link between local studies 
of planet formation and observations of disks and planets surrounding 
other nearby stars.  With an outer radius of at least 150 AU, the mass 
and size of the Kuiper Belt is comparable to the masses and sizes of 
many extrasolar debris disks \citep{bac93,art97,lag00}.  Studying planet
formation processes in the Kuiper Belt thus can yield a better 
understanding of evolutionary processes in other debris disk systems.

Making progress on planet formation in the Kuiper Belt and the dusty 
disks surrounding other stars requires plausible theories which make 
robust and testable
predictions.  This paper reviews the coagulation theory for planet 
formation in the outer solar system \citep[for reviews on other aspects 
of planet formation, see][]{man00}.  After a short summary of current 
models for planet formation, I consider recent numerical calculations 
of planet formation in the Kuiper Belt and describe observational tests 
of these models. I conclude with a discussion of future prospects for 
the calculations along with suggestions for observational tests of 
different models of planet formation.

\section{BACKGROUND}

Figure 1 shows the geometry of the outer part of our solar system.
Surrounding the Sun at the center, four colored ellipses indicate the
orbits of Jupiter (red), Saturn (cyan), Uranus (green), and Neptune
(dark blue).  The black ellipse plots the orbit of Pluto, which 
makes two orbits around the Sun for every three of Neptune.  Roughly
20\% of currently known KBOs, the {\it plutinos}, have similar orbits.
The black dots represent 200 KBOs randomly placed in the 
{\it classical Kuiper Belt,} objects in roughly circular orbits outside 
the 3:2 resonance with Neptune.  A few KBOs outside this band lie in 
the 2:1 orbital resonance with Neptune.  The eccentric magenta ellipse
indicates the orbit of one KBO in the {\it scattered Kuiper Belt}
\citep{luu97}.
The total mass in classical KBOs is $\sim$ 0.1 $M_{\oplus}$; the mass 
in scattered KBOs and KBOs in the 2:1 resonance may be comparable but is 
not so well constrained as the mass in classical KBOs \citep{tru01,gla01}.

Viewed edge-on, the orbits of the planets and the KBOs in our 
solar system lie in a disk with a height of $\sim$ 20--30 AU and a 
radius of $\sim$ 150--200 AU.  Because a disk is the natural outcome 
of the collapse of a cloud with some angular momentum, this geometry 
formed the early basis of the nebular hypothesis.  However, a cloud of 
gas and dust with the diameter of the Oort cloud, the mass of the Sun,
and a modest rotation rate of $\Omega \sim$ $10^{-8}$ yr$^{-1}$ has too 
much angular momentum to collapse to the Sun's present size.  
Building on previous realizations that a turbulent viscosity could move 
material inwards and angular momentum outwards through the protosolar 
nebula, \citet[][1948]{vW43} developed the basic physics of a viscous 
accretion disk and solved this angular momentum problem 
\citep[see also][]{lus52,ss73,lbp74}.

Most planet formation theories now begin with a viscous accretion disk
\citep[][and references therein]{kny99,man00}.  The natural evolutionary 
timescale is the viscous time scale, which measures the rate 
at which matter diffuses through the disk,
\begin{equation}
\tau_V \approx \frac{\rm 25,000~yr}{\alpha} ~ \left ( \frac{A}{\rm 100~AU} \right )^{5/4}  ~ .
\end{equation}
\noindent
This expression does not include a weak dependence on the mass of the 
central star.  The viscosity parameter $\alpha$ measures the strength of
the turbulence relative to the local thermal pressure. Most studies of 
disks in interacting binaries and other objects indicate 
$\alpha \sim 10^{-3}$ to $10^{-2}$, which yields viscous 
timescales of 1--10 Myr at 100 AU 
\citep[see][1996 for a review of the physics of accretion disks]{lp95}.  

Another evolutionary timescale for the disk depends on an external source, 
the central star, instead of internal disk physics. 
\citet[][]{hol94} showed that high energy photons from a luminous 
central star can ionize the outer skin of the gaseous disk and raise the 
gas temperature to $\sim 10^4$ K \citep[see also][1998, 2000]{shu93,ric97}.
The thermal velocity of this gas is large enough to overcome the local
gravity beyond $\sim$ 10 AU for a 1 \msun~central star.  Material then 
leaves the disk, producing a bipolar outflow which may be observed in 
nearby star-forming regions \citep{bal98,joh98}.  Disk evaporation 
occurs on a timescale 

\begin{equation}
\tau_E \approx {\rm 10^7 ~ yr} \left ( \frac{M_d}{0.01 ~ \msun} \right ) \left ( \frac{A}{\rm 10 ~ AU} \right ) \left ( \frac{\phi_{\star}}{\rm 10^{41} ~ s^{-1}} \right )^{-1/2} ~ ,
\end{equation}

\noindent
where $\phi_{\star}$ is the flux of hydrogen-ionizing photons from
the central star.

The evaporation time is sensitive to the spectral type of the central
star.  Early B-type stars with $\phi_{\star} \sim$ $10^{45}$ s$^{-1}$
can evaporate disks in $\sim$ 1 Myr or less.
The Sun has an observed flux $\phi_{\odot}$ $\sim 10^{38}$ s$^{-1}$
\citep{val81}, which leads to a long evaporation time, $\sim$ 3 Gyr, 
for a disk with $A \sim $ 100 AU.  However, young solar-type stars are 
2--3 orders of magnitude brighter than the Sun at ultraviolet and X-ray 
wavelengths \citep[e.g.,][and references therein]{dor95}.  The disk
evaporation time for a young solar-type star is therefore 
$\tau_E \approx$ 10--100 Myr 
for $A \sim$ 10--100 AU.

The evaporation and viscous timescales provide a rough upper limit to the
lifetime of a gaseous disk surrounding a solar-type star.  It is 
encouraging that both timescales are comparable to the disk lifetimes 
estimated from observations of gas and dust surrounding pre-main sequence 
stars in the solar neighborhood, $\tau_d \sim$ 1--10 Myr 
\citep{rus96,har98,lad99,bra00,hai01}.  The observational timescales place 
strong constraints on planet formation models.  Gas giants must form
before the gas disappears.  Rocky planets must form before the dust
disappears.  The observations constrain these timescales to 100 Myr
or less.

Two theories -- coagulation and dynamical instability -- try to explain 
planet formation in a viscous disk.  Coagulation theories propose that 
large dust grains decouple from the gas and settle to the midplane of 
the disk \cite{saf69,lis93}.  These grains may then 
collide to form successively larger grains \citep{wei80,wei93}
or continue to settle into a very thin layer which can become
gravitationally unstable \citep{gol73}.  Both paths produce km-sized 
planetesimals which collide and merge to produce larger bodies
\citep[][]{wei84,pal93}.  
If the growth time is short compared to the viscous timescale in the
disk, collisions and mergers eventually produce one or more `cores' 
which accumulate much of the solid mass in an annular `feeding zone' 
defined by balancing the gravity of the planetary core with the gravity of 
the Sun and the rest of the disk \citep[e.g.][and references therein]{cha01,raf01}.
Large cores with masses of 1--10 $M_{\oplus}$ accrete gas from the feeding 
zone \citep[][]{pol84,pol96,iko01}.  In our solar system, this model 
accounts for the masses of the terrestrial and several gas giant planets
\citep[][]{lis87,lis96,wei97,lev98,bry00,ida00a,ina01,ale01}.  Variants of 
this model, including orbital migration and other dynamical processes, 
can explain Jupiter-sized planets orbiting other solar-type stars 
\citep[][]{wei96,lin97,war97,for99,kle00}.  However, coagulation models 
barely succeed in making gas giant planets in 1--10 Myr, when 
observations suggest most of the gas may be gone.

Dynamical instability 
models develop the idea that part of an evolving disk can collapse 
directly into a Jupiter-mass planet \citep[e.g.,][2000]{war89,cam95,bos97}.
When the local gravity overcomes local shear and pressure forces, 
part of the disk begins to collapse.  Cool material flows into the 
growing perturbation and aids the collapse.  Eventually, the perturbation
reaches planet-sized proportions by accumulating all of the gaseous and 
solid material in the feeding zone.  This model naturally forms large 
planets on timescales, $\sim$ $10^3$ to $10^5$ yr, short compared to 
the evaporation or viscous timescales.  However, dynamical instability 
models produce neither terrestrial planets in the inner disk nor icy 
bodies like Pluto in the outer disk.  The disk mass required for a 
dynamical instability may also exceed the mass observed in pre-main 
sequence disks \citep{bec99,lad99}. 

The `Minimum Mass Solar Nebula' is an important starting point to test 
these and other planet formation models 
\citep{hoy46,wei77a,hay81,lis87,bai94}.  
The Minimum Mass is based on the near equality between the measured 
elemental compositions of the earth, moon, and meteorites 
\citep[][and references therein]{and89} and the relative abundances 
of heavy elements in the Sun \citep[see the discussion in][]{har76,ale01}. 
This analysis leads to the hypothesis that the initial elemental 
abundances of the solar nebula were nearly identical to solar abundances. 
The surface density of the Minimum Mass Solar Nebula follows from 
adding hydrogen and helium to each planet to reach a solar abundance 
and spreading the resulting mass uniformly over an annulus centered 
on the present orbit of the planet.

Figure 2 shows how the surface mass density varies with distance 
for the Minimum Mass Solar Nebula.  The arrows indicate the mass
added to the terrestrial planets.  The plot shows Venus, Earth, 
Jupiter, Saturn, Uranus, Neptune, and the Kuiper Belt.  When the 
material at the orbits of Venus and Earth is augmented to reach a 
solar abundance of hydrogen, the surface density for the gas follows 
the solid curve, 
$\Sigma_g \approx ~ \Sigma_0 ~ (A / {\rm 1 ~ AU}) ^{-3/2}$, out to 
$A \approx$ 10 AU and then decreases sharply.  The solid curve
in Figure 2 has $\Sigma_0$ = 1500 g cm$^{-2}$; for comparison,
\citet{hay85} concluded $\Sigma_0$ = 1700 g cm$^{-2}$ while
\citet{wei77a} proposed $\Sigma_0$ = 3200 g cm$^{-2}$.
Following \citet{hay81}, the dot-dashed curve representing
the mass density of solid material has

\begin{equation}
\Sigma_s = \left\{ \begin{array}{l l l}
	   7 ~ {\rm g ~ cm^{-2}} ~ (A / {\rm 1 ~ AU}) ^{-3/2} & \hspace{5mm} & A \le 2.7 ~ {\rm AU} \\
                     \\
	   30 ~ {\rm g ~ cm^{-2}} ~ (A / {\rm 1 ~ AU}) ^{-3/2} & \hspace{5mm} & A > 2.7 ~ {\rm AU} \\
 \end{array}
         \right .
\end{equation}

\vskip 2ex
\noindent
The uncertainties in the coefficients are a factor of $\sim$ 2.  The 
change in the surface density of solid material at 2.7 AU corresponds to 
the region where ice condenses out of the gas in the \citet{hay81} model.
The location of this region depends on the disk structure 
\citep{sas00}.

The Minimum Mass Solar Nebula was one of the great successes of
early viscous accretion disk theories, because steady-state disk models 
often yield $\Sigma \propto A^{-3/2}$.  The sharp decrease in the 
``observed'' $\Sigma$ at 10--30 AU supports photoevaporation models where 
ionized hydrogen becomes unbound at $\sim$ 10 AU \citep{shu93}.
Current abundance measurements for the gas giants lend additional 
evidence: the gas-to-dust ratio appears to decrease with heliocentric 
distance in parallel with the surface density drop beyond 10 AU
\citep{pol84,pod85,pod87,pol96}.  In the Kuiper Belt, there may be 
two origins for the large drop in the observed surface density from a 
$\Sigma \propto A^{-3/2}$ model.  Adding H and He to achieve a solar
abundance at 30--40 AU increases the mass in the Kuiper Belt by 
a factor of $\sim$ 30. Material lost to orbital dynamics and to
high velocity collisions of objects in the Belt may increase the 
current mass by another factor of 10--100 \citep[e.g.][]{hol93,dav97},
bringing the
initial surface density in the Kuiper Belt within range of the
$\Sigma \propto A^{-3/2}$ line.  If these estimates are correct,
the total mass of the Minimum Mass Solar Nebula is $\sim$ 0.01 \msun~for
an outer radius of $\sim$ 100 AU, close to the median mass for 
circumstellar disks surrounding young stars in nearby regions 
of star formation \citep{lad99}.

Figure 2 suggests that the Kuiper Belt provides an important test 
of coagulation models.  Forming objects with radii of $\sim$ 
500--1000 km requires $\sim$ 10--100 Myr at $\sim$ 40 AU in a Minimum 
Mass Solar Nebula (see below).  The outermost gas giant, Neptune,
must form on a similar timescale to accrete gas from the solar nebula 
before the gas escapes (equations (1--2)).  Neptune formation places 
another constraint on the KBO growth time; Neptune inhibits 
KBO formation at 30--40 AU by increasing particle random velocities 
on timescales of 20--100 Myr \citep{hol93,lev93,dun95,mal96,mor97}.
\citet[][1999a,b]{kl98} investigated how KBOs form by coagulation 
and compared their results with observations 
\citep[see also][1997b]{fer97,st97a}.
The next section briefly describes the model results; \S4 
compares these results with observations.

\section{Kuiper Belt Models}

\subsection{Coagulation Calculations}

\citet{saf69} invented the current approach to planetesimal accretion
calculations.  In his particle-in-a-box method, planetesimals are a 
statistical ensemble of masses with a distribution of horizontal and 
vertical velocities about a Keplerian orbit 
\citep[see also][]{gre78,oht88b,wet89,spa91,ste95,kl98}.  
Because $n$-body codes cannot follow the $10^{15}$ or more small 
planetesimals required in a typical coagulation calculation, the 
statistical approximation is essential. The model provides a kinetic 
description of the collision rate in terms of the number density and 
the gravitational cross-section of each type of planetesimal in the grid.
Treating planetesimal velocities as perturbations about a Keplerian 
orbit allows the use of the Fokker-Planck equation to solve for changes in 
the velocities due to gravitational interactions and physical collisions.  

In our implementation of Safronov's model, we begin with a differential 
mass distribution, $n(m_i$), in concentric annuli centered at 
heliocentric distances, $A_j$, from a star of mass $M_{\star}$ 
\citep[][2002]{kl99a,kb01}.  The mass distribution has $N$ mass 
batches in each annulus; $\delta_i \equiv m_{i+1} / m_i$ is the 
mass spacing between batches.  To provide good estimates of the 
growth time, our calculations have $\delta$ = 1.1--2.0 
\citep{oht88a,wet90,kol92,kl98}.  To evolve the mass and 
velocity distributions in time, we solve the coagulation 
and energy conservation equations for an ensemble of objects with 
masses ranging from $\sim 10^{7}$~g to $\sim 10^{26}$ g.
We adopt analytic cross-sections to derive collision rates, use the
center-of-mass collision energy to infer the collision outcome
(merger, merger + debris, rebound, or disruption), and compute 
velocity changes from gas drag \citep{ada76,wei77b,wet93},
Poynting-Robertson drag \citep{bur79,kar93}, and collective 
interactions such as dynamical friction and viscous stirring using 
a Fokker-Planck integrator \citep{ste80,hor85,bar90,wet93,luc95,oht99,ste00}.  
The code reproduces previous calculations for accretion at 
1 AU \citep{wet93,wei97}, collisional disruption of pre-existing 
large KBOs at 40 AU \citep{dav97}, and $n$-body simulations of 
gravitational scattering at 1 AU \citep[see][and references therein]{kb01}.

During the early stages of planet formation, particle-in-a-box algorithms 
yield good solutions to the coagulation equation 
\citep{ida92,kok96,lee00,mal01}.  Most published calculations have been 
made for a single accumulation zone to get a good understanding of the 
basic physics without spending a large amount of computer time 
\citep[e.g.,][]{gre78,oht88b,wet89,ste96a,kl98}.  
Single annulus calculations provide the basis for estimates of planetary 
growth rates as a function of heliocentric distance and initial disk mass 
\citep{lis87,wet93,lis96}.  Multiannulus calculations allow bodies in
neighboring accumulation zones to interact and thus provide better
estimates of planetary growth rates \citep{spa91,kb02}.  These codes 
enable calculations
with additional physics, such as orbital migration, which cannot be
incorporated accurately into single annulus codes. Once large objects 
form, one-on-one collisions become important; statistical estimates 
for collision cross-sections and gravitational stirring in single and
multiannulus codes begin to fail.  More detailed $n$-body calculations 
are then required to study the evolution of the largest 
objects.

In the following sections, I discuss published single annulus models
for Kuiper Belt objects and then outline new multiannulus calculations.

\subsection{Single Annulus Models}

Our Kuiper Belt models begin with an input cumulative size distribution 
\begin{equation}
N_C \propto r_i^{-q_i}, 
\end{equation}
\noindent
with initial radii $r_i$ = 1--80 m and $q_i$ = 3.  These particles 
are uniformly distributed in a single annulus with a width of 6 AU 
at 32--38 AU from the Sun.  The total mass in the annulus is $M_0$; 
$M_0 \approx$ 10 $M_{\oplus}$ for a Minimum Mass Solar Nebula.  All mass 
batches start with the same initial eccentricity $e_0$ and inclination
$i_0 = e_0/2$.  The adopted mass density, $\rho_0$ = 1.5 g cm$^{-3}$, 
is appropriate for icy bodies with a small rocky component
\citep{gre98}.  These
bodies have an intrinsic tensile strength $S_0$ which is independent
of particle size and a total strength equal to the sum of $S_0$ and 
the gravitational binding energy \citep[][1994]{dav85}.  
\citet[][1999b]{kl99a} describe these parameters in more detail.

Figure 3 shows the results of a complete coagulation calculation for 
$M_0$ = 10 $M_{\oplus}$, $e_0$ = $10^{-3}$, and $S_0$ = $2 \times 10^6$
erg g$^{-1}$ \citep[see also][1997b]{ste96a,st97a}.  We separate the 
growth of KBOs into three regimes.  Early in the evolution, frequent 
collisions damp the velocity dispersions of small bodies.  Rapid growth
of these bodies erases many of the initial conditions, including $q_i$ 
and $e_0$ \citep[][1999]{kl98}. These bodies slowly grow into 1~km 
objects on a timescale of 5--10 Myr $(M_0/10 ~ {\rm M_{\oplus}} )^{-1}$.  
The timescale is sensitive to the initial range of sizes; because 
collisional damping is important, models starting with larger objects 
take longer to reach runaway growth.  The linear growth phase ends when 
the gravitational range of the largest objects exceeds their geometric 
cross-section.  Gravitational focusing enhances the collision rate by 
factors of $(V_e/V_c)^2 \approx$ 10--1000, where $V_c$ is the collision
velocity and $V_e$ is the escape velocity of a merged object.  The 
largest objects then begin ``runaway growth'' \citep[e.g.,][]{gre78,wet93}, 
where their radii grow 
from $\sim$ 1~km to $\gtrsim$ 100~km in several Myr.  During this phase, 
dynamical friction and viscous stirring increase the velocity dispersions 
of the smallest bodies from $\sim$ 1~m~s$^{-1}$ up to $\sim$ 40~m~s$^{-1}$.
This velocity evolution reduces gravitational focusing factors and ends 
runaway growth.  The largest objects then grow slowly to 1000+ km sizes 
on timescales that again depend on the initial mass in the annulus.
\citet{kok98} call this last phase in the evolution `oligarchic growth' 
to distinguish it from the linear and runaway growth phases 
\citep[see also][]{ida93}.

The shapes of the curves in Figure 3 show features common to all
coagulation calculations \citep[e.g.,][Davis, Farinella, \&
Weidenschilling 1999]{wet89,st97a,st97b,wei97}.  
Almost all codes produce two power-law size distributions. The merger 
component at large sizes has $q_f \approx$ 3; the debris component 
at small sizes has $q_f = 2.5$ \citep{doh69,tan96}.  
Dynamical friction produces a power law velocity distribution 
in the merger component.  The debris component has roughly constant
velocity, because it contains a small fraction of the initial mass.
The transition region between the two components 
usually has a `bump' in the size distribution, where objects which 
can merge grow rapidly to join the merger population \citep{dav97,dav99}.
Calculations for annuli closer to the Sun also yield a `runaway' population, 
a plateau in the size distribution of the largest objects \citep{wet93}.
The objects in this plateau contain most of the mass remaining in the
annulus \citep{wet93,wei97}.
In our models, the largest 10--20 objects are not massive enough to 
produce a `runaway plateau' in the size distribution until the very
late stages of the evolution (see below).

Our Kuiper Belt calculations yield one result which is very different
from coagulation calculations for annuli at less than 10 AU from the
Sun.  In all other published calculations, the largest bodies contain
most of the initial mass in the annulus.  In the Kuiper Belt, most of
the initial mass ends up in 1 km objects.  Fragmentation and gravitational
stirring are responsible for this difference between calculations at
1--10 AU and at 40 AU.  In our calculations, fragmentation produces a 
large reservoir of small bodies that damp the velocity dispersions of 
the large objects through dynamical friction.  These processes allow 
a short runaway growth phase where 1 km objects grow into 100 km objects.  
Continued fragmentation and velocity evolution damp runaway growth by 
increasing the velocity dispersions of small objects and reducing
gravitational focusing factors.  Our models thus
enter the phase of `oligarchic growth' earlier than models for planet 
growth at 1--10 AU.  This evolution leaves $\sim$ 1\%--2\% of the 
initial mass in 100--1000 km objects.  The remaining mass is in 
0.1--10~km radius objects.  Continued fragmentation gradually 
erodes these smaller objects into dust grains that are removed from 
the Kuiper Belt on short timescales, $\sim 10^7$ yr 
\citep{bac93,bac95,dav97,dav99}.  Thus, in our interpretation, 
100--1000 km radius objects comprise a small fraction of the original 
Kuiper Belt.  

Planet formation in the outer parts of a solar system is self-limiting.
During the late stages of planetesimal evolution, large planets stir 
smaller objects up to the shattering velocity.  This process leads to
a collisional cascade, where planetesimals are ground down into smaller
and smaller objects.  Continued fragmentation, radiation pressure, and 
Poynting-Robertson drag then remove small particles from the disk 
faster than large objects can accrete.  Because the shattering velocity
depends on the tensile strength of a planetesimal, collisional cascades 
start sooner when planetesimals are weaker.  The maximum mass of 
an icy object in the outer solar system thus depends on its strength 
(Figure 4).  At 35--140 AU, our calculations yield a linear relation 
between the maximum radius and the intrinsic strength of a 
planetesimal,
\begin{equation}
{\rm log} ~ r_{max} \approx 2.45 - 0.09 ~ {\rm log} (a_i / {\rm 35 ~ AU}) + 0.22 ~ {\rm log} ~ S_0 ~ ,
\end{equation}
for planetesimals with log $S_0$ = 1--6 \citep{kl99a}.

The weak variation of $r_{max}$ with heliocentric distance is a new result
based on calculations for this review.  If planetesimals all have the
same strength, the shattering velocity is independent of heliocentric 
distance.  Once small planetesimals reach the shattering velocity, the 
largest objects do not grow.  Because planetesimals at larger heliocentric 
distances are less bound to the central star, a massive planet at large 
$a_i$ stirs small planetesimals more effectively than the same planet 
in orbit at small $a_i$.  Small planetesimals in the outer part of the 
disk thus require relatively less stirring to reach the shattering 
velocity than small planetesimals in the inner part of the disk.  A 
less massive planet in the outer disk can stir planetesimals to the 
shattering limit as effectively as a more massive planet in the inner 
disk.  Hence, larger objects form in the inner disk than in the outer disk.

The initial mass $M_0$ is the main input parameter which establishes
the formation timescale and the mass distribution of KBOs in the outer
solar system.  Figure 5 illustrates the time variation of the model 
parameter $r_5$, defined as the radius where the cumulative number of 
objects exceeds $10^5$ \citep{kl99a}.  Most surveys estimate $\sim$
$10^5$ KBOs with radii of 50 km or larger; the $r_5$ parameter thus provides 
a convenient way to compare theory with observations.  Figure 5 shows 
that $r_5$ increases steadily with time during the linear growth phase.
The number of 50 km radius KBOs 
increases dramatically during runaway growth and then approaches a 
roughly constant value during oligarchic growth.  More massive models
enter runaway growth sooner; the timescale for $r_5$ to reach 50 km
is 
\begin{equation}
\tau(r_5 = 50~{\rm km}) \approx 10 {\rm ~ Myr ~ } (M_0/10~M_{\oplus})^{-1}
\end{equation}
More massive disks also produce more 50 km radius KBOs.  Based on
Figure 5, protosolar nebulae with less than 30\% of the Minimum Mass
produce too few 50 km radius KBOs; nebulae with more than $\sim$ 3
times the Minimum Mass may produce too many.

The timescale for Pluto formation at 35 AU is also sensitive to the 
initial population of bodies with radii of 1 km or smaller.
Collisional damping of these small bodies leads to an early runaway 
growth phase where 0.1--1 km bodies grow rapidly to sizes of 100 km or
larger.  Because collisional damping is ineffective for bodies with radii 
1--10 km or larger, calculations which exclude small bodies take at 
least a factor of 3 longer to reach runaway growth \citep{kl98}.  
These models also fail to achieve a shallow power-law size distribution
with $q_f$ = 3 until late in the oligarchic growth phase 
\citep[e.g.,][]{dav99}.  

Pluto formation is remarkably insensitive to other initial conditions 
in the disk.  Growth by mergers, collisional damping, and dynamical 
friction rapidly erase the initial size and velocity distributions.  As 
long as particle strengths exceed a minimum value of 300 erg g$^{-1}$,
the details of the fragmentation algorithm do not affect planetesimal 
growth significantly.  Formation times change by a factor of two or less 
for order of magnitude changes in the fragmentation parameters and the
initial size and velocity distributions \citep[][1999b]{kl99a}.

\subsection{Multiannulus calculations}

Multiannulus calculations address some of the limitations and uncertainties
of coagulation models in a single accumulation zone \citep{spa91,wei97}.
By including long-range interactions between objects in neighboring annuli,
a multiannulus code yields better treatment of velocity evolution and
more accurate estimates for the accretion rates of large bodies.  The
improvement resulting from a multiannulus code scales with the number 
of annuli.  More annuli allow a more accurate treatment of collision 
cross-sections and velocity evolution \citep[][2002]{kb01}.

To illustrate some results from our multiannulus code, I describe
two calculations of large planetesimals in the Kuiper Belt.  The
calculations begin with 0.1--1.0 km objects in 16 annuli at distances 
of 40--54 AU from the Sun.  The planetesimals have an initial 
eccentricity $e_0 = 2 \times 10^{-3}$ and a tensile strength
$S_0$ = $2 \times 10^6$ erg g$^{-1}$.  The debris receives a small 
fraction, $f_{KE}$ = 0.05, of the impact kinetic energy. The 
calculations do not 
include gas drag or Poynting-Robertson drag.  During the 1--5 Gyr of 
each calculation, drag forces have negligible impact on the evolution 
of objects with radii of 0.1 km or larger.

Figure 6 illustrates the time evolution of the size and horizontal
velocity distributions for a model with fragmentation \citep{gre84,dav85,kl99a}
and velocity evolution \citep{ste00,kb01}.  During the first 20 Myr 
of this calculation, collisions damp the velocity dispersions of the 
smallest bodies.  Planetesimals grow slowly from 1 km to $\sim$ 10 km.  
When objects are larger than $\sim$ 10 km, gravitational focusing 
enhances collision rates.  The largest objects then grow rapidly to 
sizes of $\sim$ 200--300 km.  Dynamical friction and viscous stirring 
heat up the orbits of the smallest objects.  This evolution reduces 
gravitational focusing factors and ends runaway growth.  A handful of 
large objects then grow slowly; their sizes reach $\sim$ 1000 km at 
70 Myr and $\sim$ 3000 km at $\sim$ 120 Myr.  

The lower panel of Figure 7 illustrates the evolution of the largest
body in each annulus.  Collisions are most rapid in the inner annuli; 
objects at 40 AU thus grow faster than objects at 50 AU.  Runaway
growth begins first at 40 AU (10--20 Myr), when objects at 55 AU 
have grown by less than a factor of two.  After 30 Myr, the largest 
objects at 40 AU have radii of $\sim$ 100 km and then grow slowly to
radii of $\sim$ 1000 km during the oligarchic growth phase.  This
evolution is delayed at 50 AU.  During runaway growth, objects at 
50--55 AU grow from sizes of $\sim$ 10 km at 30--50 Myr to $\sim$ 
100--200 km at 50--70 Myr.  After $\sim$ 100 Myr, the largest objects
in all annuli grow slowly at roughly the same pace.

Objects grow much more slowly in models without fragmentation and 
velocity evolution (Figure 7, top panel).  During the first 300 Myr 
of the calculation, planetesimals grow slowly from $\sim$ 1 km to 
$\sim$ 10 km.  Because particle velocities are constant, gravitational 
focusing factors change little.  Once particle sizes reach $\sim$ 100 km, 
runaway growth begins in the innermost two annuli.  
A few large bodies rapidly accrete most of the mass in each annulus.  
At $\sim$ 500 Myr, the largest body in the second annulus accretes 
the largest body in the first annulus and then consumes the rest of 
the bodies in annuli 1--5. Large bodies in annuli 6--8 begin runaway 
growth at $\sim$ 700 Myr.  A single large body in annulus 7 consumes 
all of the bodies in annuli 6--10.  This process repeats for annuli 
11--13 at $\sim$ 1 Gyr, when a single large body in annulus 11 grows
almost as large as the bodies in annuli 2 and 7.  The remaining objects 
in annuli 14--16 probably form a fourth large object at $\sim$ 1.3 Gyr;
we terminated the calculation before this point.

Figure 8 compares the evolution of the largest bodies in each calculation.
Collisional damping dominates the velocity evolution of small particles
at 40--55 AU \citep[see also][1999a]{kl98}.  Dynamical friction provides
additional damping to the largest bodies.  Smaller particle velocities
produce larger gravitational focusing factors and more rapid growth rates.
Models with velocity evolution thus enter the runaway growth phase 
earlier ($\sim$ 10--30 Myr) than models without velocity evolution
($\sim$ 300--500 Myr).  During runaway growth, viscous stirring 
dominates the velocity evolution of all particles.  Larger particle
velocities yield smaller gravitational focusing factors and smaller
growth rates.  By removing small particles from the grid, fragmentation 
reduces growth rates further.  Thus, the largest bodies reach a 
maximum size which depends on the strength of the smallest bodies and 
the heliocentric distance \citep[Figure 4; see also][]{kl99a}.
In models without velocity evolution, gravitational
focusing factors grow with the mass of the largest body.  Thus, 
models without velocity evolution produce a few very massive objects.
The orbital separation of these massive objects is roughly their
gravitational range.  In our calculations, this limit is 
$\sim$ 2.4 Hill radii; $R_H = (m_p/3~M_{\odot})^{1/3}$, where
$m_p$ is the mass of the planet.

The results for KBO formation in these initial multiannulus calculations
are encouraging.  Successful KBO models need to form $\sim 10^5$ KBOs 
and at least one Pluto before Neptune attains its present mass at $\sim$ 
25 AU \citep[e.g.,][1999a]{kl98}.  If the gas in the solar nebula is 
depleted on timescales of 5--10 Myr, Neptune must form on similar 
timescales \citep{bry00}.  Some recent numerical calculations of gas 
accretion onto rocky cores can achieve this goal
\citep{fer84,ip89,pol96,bry00}.  Although our Pluto formation timescale
of 60--70 Myr is long compared to these constraints, single annulus
calculations starting from smaller bodies, 1--100 m in size, form 
Pluto and numerous KBOs on timescales of 10--20 Myr 
\citep[Figures 3 and 5;][]{kl99a}.  Scaling the single annulus 
models suggests formation timescales of 5--20 Myr at 40--50 AU 
with a multiannulus code. 

\subsection{Long term evolution}

Several processes shape the long-term evolution of KBOs in the outer solar 
system \citep{hol93,bac95,dav97,tep99,dav99,kuc02}.  Once Neptune attains 
its current mass and location, gravitational perturbations pump up orbital 
velocities and begin to remove KBOs of all sizes from the Kuiper Belt.  
Gravitational stirring by the largest KBOs increases orbital velocities 
of smaller KBOs to the shattering limit.  Once a collisional cascade 
begins, the largest objects do not grow significantly.  Small objects are 
shattered and then removed from the Kuiper Belt by radiation pressure and 
Poynting-Robertson drag.  

To begin to understand how these processes have shaped the current
population of KBOs, several groups have calculated the long-term collisional 
evolution of large objects in the Kuiper Belt.  \citet{dav97} used a single
annulus code to show that the observed population of KBOs with radii of 
25--50 km or larger can survive disruptive collisions for 5 Gyr at 
40--50 AU \citep[][1997b]{ste96b,st97a}.  
These objects are thus remnants of the original population formed during
the early evolution of the Kuiper Belt.  For reasonable values of $S_0$,
smaller KBOs are collision fragments produced during the collisional 
cascade.  \citet{dav99} confirmed these results.  For $t > 1$ Gyr, the 
\citet{dav99} calculations yield a very steep power law size distribution 
for the merger population, $q_f \approx$ 11. This result differs from the 
results of single annulus codes and the multiannulus result in Figure 6.
The source of this difference is uncertain but may be due to different 
treatments of velocity evolution or fragmentation.  

Figures 9--10 illustrate how the size distribution evolves at late times
in our multiannulus calculations.  The first model is a continuation of
the calculation for Fig. 6; the second model repeats this calculation
for weak bodies with $S_0$ = $10^3$ erg g$^{-1}$.  The first 70--100 Myr 
of this second calculation closely follows the evolution of the first 
model. Runaway growth at 40--50 AU produces 10 km bodies in 20 Myr and
100 km bodies in 30 Myr.  Slower oligarchic growth leads to 1000 km 
bodies at 70 Myr (40 AU) to 100 Myr (50 AU).  At 70 Myr, the amount of
debris produced from collisional erosion of the small bodies is 
negligible, $\sim$ 6\% of the mass at the start of the calculation.

Models with strong, icy particles (Fig. 9; $S_0$ = $2 \times 10^6$ erg 
g$^{-1}$) have a long oligarchic growth phase followed by a 
collisional cascade.  The largest objects grow from $r_i \sim$ 1000 km 
at $t$ = 70 Myr to $r_i \sim$ 3000 km at $t = 300$ Myr to 
$r_i \sim$ 6000 km at $t = 1$ Gyr.  This slow growth phase produces 
a power-law size distribution, with $q_f \approx 3.35$ for $r_i$ = 
20--6000 km.  As the largest bodies reach sizes of 2000--6000 km,
viscous stirring slowly increases 
the eccentricities of the smallest objects from $e \approx$ 0.01 at 
70 Myr to $e \approx$ 0.05 at 300 Myr to $e \approx$ 0.09 at $t$ = 1 Gyr.  
Throughout most of this phase, collisions between small objects produce 
debris through cratering; this debris is $\sim$ 10\% of the initial mass 
at 300 Myr and $\sim$ 70\% of the initial mass at 1 Gyr.  Cratering
removes the bump in the size distribution for $r_i \sim$ 1 km.  Just before 
1 Gyr, collisional disruption begins to deplete the population of 0.1--1 km 
bodies on timescales of $\sim$ 100 Myr.  This evolution starts to 
produce a dip in the size distribution at $r_i \approx$ 1 km.

Collisional cascades begin sooner in models with weak bodies (Fig. 10; 
$S_0$ = $10^3$ erg g$^{-1}$).  Cratering is not important in these models;
most mass is lost through collisional disruption of small bodies.  
During the first 70 Myr of evolution, cratering is responsible for 
less than 1\% of the mass loss; collisional disruption removes $\sim$ 
6\% of the initial mass.  After 70 Myr, objects grow slowly as more
and more material is lost to collisional disruption.  The largest
object has a radius of $\sim$ 2000 km at 300 Myr and $\sim$ 3000 km
at 1 Gyr.  The size distribution for the largest bodies follows a power 
law with $q_f \approx 3.15$ for $r_i$ = 20 km to 2000--3000 km.
For smaller bodies, collisional disruption produces a pronounced dip
in the size distribution at 0.3--1.0 km.  The debris lost to bodies
with $r_i <$ 0.1 km (the smallest object in the grid) is
$\sim$ 65\% of the initial mass at 300 Myr and
$\sim$ 85\% of the initial mass at 1 Gyr.

These multiannulus calculations confirm some of the single annulus
results.  The size of the largest object at 40--50 AU depends on the
tensile strength of 0.1--10 km objects.  Stronger small bodies allow
the growth of larger large bodies \citep[Figure 4;][]{kl99a}.  We plan 
additional calculations to see whether the size-strength relation is similar 
to equation (5); preliminary results suggest a shallower relation.  
Gravitational stirring by 1000 km and larger objects in the grid leads 
to a collisional cascade, where cratering and collisional disruption
remove small bodies from the grid \citep[see also][1997b]{dav97,dav99,st97a}.
The duration of the collisional cascade is $\sim$ 100 Myr to $\sim$
1 Gyr \citep[see also][]{kb01}.  Collisions convert $\sim$ 80\% to 90\%
of the initial mass in the grid to small particles with sizes of 
100 m or smaller.  Disruptive collisions and Poynting-Robertson drag
can remove this material from the Kuiper Belt on short timescales,
$\sim$ 10--100 Myr \citep[Backman et al. 1995;][]{ste96b,tep99}.

During the late stages of our multiannulus calculations, the size 
distribution for the largest objects follows a power law with 
$q_f$ = 3.15--3.35.  Once the largest objects have radii of $\sim$
1000 km or larger, the slope of the power law size distribution is
nearly invariant.  We plan additional calculations to test the 
sensitivity of the slope to initial conditions and the fragmentation 
parameters.

\section{OBSERVATIONAL TESTS OF COAGULATION MODELS}

Observations provide powerful constraints on the KBO population.  
Sensitive imagers on large ground-based and space-based telescopes 
detect individual large objects directly.
Current instrumentation yields direct detections of 50 km objects 
from the ground and 10 km objects from the {\it Hubble Space Telescope
(HST)}.  Future large ground-based 30-m to 100-m telescopes and the
{\it Next Generation Space Telescope (NGST)} will improve these 
limits by an order of magnitude or more.  The population of smaller 
KBOs with radii of $\sim$ 1 km can be estimated indirectly from the
frequency of short-period comets and from dynamical calculations.
The population of KBOs with sizes smaller than $\sim$ 0.1 km can 
only be derived as an ensemble by measuring the surface brightness 
of the sky and eliminating other radiation sources.  Despite confusion 
from the galaxy and the local Zodiac, optical and far-infrared (far-IR) 
observations provide useful measures of the population of dust grains 
in the Kuiper Belt.

These data allow broad observational tests of KBO formation models.
The large sample of individual KBO detections provides a good measurement
of KBO number counts, the number of KBOs per magnitude per square degree 
projected on the sky.  For an adopted albedo $\omega_l$ for large KBOs, 
the number counts directly yield the KBO size distribution for objects 
with radii of 50 km or larger \citep{jew98,luu98,chi99,gla01}.   
The radial distribution of large 
KBOs follows from the size distribution and heliocentric distances derived
from the orbit or from an adopted albedo \citep{don97,all01,tr01,tru01}.
Surface brightness measurements constrain the size distribution of
small KBOs.  Far-IR data measure thermal emission from small grains
in the Kuiper Belt; optical and near-IR data measure scattered light.
Deriving constraints on the size distribution from surface brightness
data requires an assumption about the grain albedo $\omega_g$, which
may differ from $\omega_l$.

\subsection{Number Counts}

The observed number counts of bright KBOs follow a simple relation
\begin{equation}
{\rm log} ~ N = \alpha (R - R_0) ~ ,
\end{equation}
\noindent
where $N$ is the cumulative number of bodies brighter than magnitude
$R$ \citep[][]{gla98,jew98,chi99}.  
Recent fits to the observations suggest $\alpha$ = 0.65--0.70 and
$R_0$ = 23.3--23.5 \citep{gla01,tru01b}.
If the size distribution of KBOs is independent of heliocentric distance 
and if all KBOs have the same albedo, a power-law relation for the 
number counts implies a power-law size distribution,
\begin{equation}
{\rm log} ~ N_C = N_0 \left ( \frac{r}{r_0} \right )^{-q} ~ ,
\end{equation}
\noindent
where $N_C$ is the cumulative number of objects with radius
larger than $r$ and $q$ = 5$\alpha$ \citep{jew98,chi99,gla01}.  
Fits to the observations thus imply size distributions with
$q$ = 3.25--3.50.
The characteristic radius $r_0$ is related to $R_0$; the scaling factor
$N_0$ depends on the total mass in the Kuiper Belt.  

\citet{kl99b} show that complete coagulation calculations produce power 
law size distributions for large KBOs.  For a wide range of input parameters, 
single annulus models yield $q$ = 2.75--3.25 for KBOs with radii of 
$\sim$ 10--1000 km \citep[see Table 2 of][]{kl99b}.  To construct predicted 
number counts, \citet{kl99b} adopt $\omega_l$ = 0.04 and the slope parameter,
$g$ = 0.15, in the standard two parameter magnitude relation for asteroids 
\citep{bow89}.  An adopted heliocentric distance $d$ and a random phase angle 
$\beta$ from the Sun then specify the observed R magnitude for a KBO with 
radius $r_i$.  The slope parameter $g$ relates the brightness of an asteroid 
at solar phase angle $\beta$ to the brightness at opposition, $\beta$ = 
0\degree.  \citet{kl99b} assume that the KBO size distribution is 
independent of heliocentric distance, with 50\% of the KBOs in a ring 
at 42--50 AU and the rest as Plutinos at 39.4 $\pm$ 0.2 AU.  The resulting 
number counts are insensitive to the Plutino fraction and the outer radius 
of the ring.

The upper panel of Figure 11 compares predicted with observed number counts 
from several single annulus calculations.  Data are as indicated in the 
legend.  Error bars for the measured points are typically a factor of 2--3
and are not shown for clarity.  The lines plot predicted number counts 
for models with $e_0 = 10^{-3}$ and $M_0 \approx$ 0.3 (dot-dashed),
1.0 (solid), and 3.0 (dashed) times the Minimum Mass Solar Nebula.  
Models with different $e_0$ are indistinguishable for $R \le$ 27
\citep{kl99b}.  The model luminosity functions agree well with current 
observations.

Multiannulus calculations also produce power-law size distributions for
large KBOs (Figures 6 and 9--10).  For several completed calculations, 
these models yield steeper slopes, $q$ = 3.2--3.5, for the size 
distribution of objects with radii of 10--1000 km.  These results are
much closer to the observed slopes than the multiannulus calculations
of \citet{dav99}.  We plan additional 
multiannulus calculations to measure the scatter in 
the predicted slope of the size distribution.  To construct an initial model 
for the number counts, I use radial distributions of KBOs derived from 
the coagulation code and adopt $e$ = 0 and $\beta$ = 0 for all sources.  
This model assumes all sources are found at opposition and neglects 
bright KBOs closer than 40 AU.

The lower panel of Figure 11 compares observed number counts with 
predictions for several multiannulus calculations.  The data are the 
same in both panels.  The lines show predicted
number counts for multiannulus models with an initial mass in solid
material equal to the Minimum Mass Solar Nebula.  The solid curve 
indicates counts when the first Plutos form at 40--45 AU.  The other 
curves plot counts at 1 Gyr for models where the tensile strength of 
small objects is $S_0 = 10^3$ erg g$^{-1}$ (dot-dashed curve) and
$S_0 = 2 \times 10^6$ erg g$^{-1}$ (dashed curve).  For $R \ge$ 20,
model counts at 1 Gyr are independent of $S_0$.  Models with stronger
planetesimals produce larger planets and thus predict more objects 
with $R \le$ 20 at 1 Gyr. 

The good agreement between models and observations for R = 20--26 in
Figure 11 is encouraging.  When the first Plutos form at 40--45 AU
in the multi-annulus calculations,
the predicted number counts follow a linear relation between log $N$
and $R$ (equation (7)) with $\alpha$ = 0.80$\pm$0.01 and $R_0$ = 
22.45$\pm$0.05.  After 1 Gyr, the slope of the number counts is
$\alpha$ = 0.65$\pm$ 0.02, much closer to the value derived from
the data, $\alpha$ = 0.65--0.70 \citep{gla01}.  The normalization 
derived for the models, $R_0$ = 21.95$\pm$0.10, is roughly a magnitude
larger than the measured $R_0$ = 23.0--23.5.  However, these models do
not include loss of KBOs by dynamical interactions with Neptune. 
At 40--50 AU, these dynamical losses range from $\sim$ 50\% to $\sim$ 
80\% of the initial mass in the Kuiper Belt \citep[e.g.,][]{hol93,lev95}.
Applying these losses to our 1 Gyr number count models yields
$R_0 \sim$ 22.70--23.70, passably close to the observed value.

\subsection{KBOs and Olbers Paradox}

Many KBOs are too faint to be detected as individual objects even with 
large telescopes.  All together, these faint KBOs can produce a detectable 
diffuse background light.  Optical and near-IR data measure the amount of 
scattered light from faint KBOs; far-IR and submm data measure the amount 
of thermal emission.  The KBO background light is smaller than diffuse 
emission from the local Zodiac \citep{lei98} and has not been detected 
\citep[Backman et al. 1995;][]{ste96b,tep99}.  Nevertheless, the upper limits
on scattered and thermal emission provide interesting constraints on
the population of small KBOs.

Measured optical and far-IR sky surface brightnesses demonstrate that
KBO number counts cannot follow equation (7) to arbitrarily faint 
magnitudes.  For equation (7) with $\alpha > 0.4$, the optical sky 
surface brightness of KBOs brighter than magnitude $R$ is \citep{kw01}:
\begin{equation}
\mu_R = 41.03 - 2.5 ~ {\rm log} \left ( \frac{\alpha}{\alpha - 0.4} \right ) + (1 - 2.5 \alpha) ~ (R - R_0) ~ .
\end{equation}
\noindent
This surface brightness exceeds the measured sky surface brightness in the 
ecliptic plane,\footnote{The observed flux of the Zodiacal light decreases 
away from the ecliptic plane as csc $\beta$ where $\beta$ is the ecliptic 
latitude. Using the measured surface brightness at $\beta$ = 30\degree, 
the approximate vertical thickness of the KBO distribution, does not change 
the main conclusions of this section.} $\mu_R \approx$ 22 mag arcsec$^{-2}$,
at $R \approx$ 
45--55 for $\alpha \approx$ 0.6--0.75 \citep{win92,win94,win98,bir00,kw01}.
For an adopted albedo $\omega_g$ and temperature $T_{KBO}$, the thermal 
background from small KBOs depends only on the optical surface brightness
\begin{equation}
I_{\nu} ({\rm FIR}) \lesssim 9.5 \times 10^{17 - 0.4\mu_R} ~ T_{KBO}^{-1} ~ \left ( \frac{1 - \omega_g}{\omega_g} \right )~ {\rm Jy ~ sr^{-1}} ~ .
\end{equation}
\noindent
This result assumes that a small KBO emits less radiation than the maximum 
flux of a blackbody with temperature $T_{KBO}$.  For $\omega_g \approx$ 0.5, 
small KBOs with $\mu_R \gtrsim$ 22 mag arcsec$^{-2}$ and 
$T_{KBO} \approx$ 40 K \citep{bac95,tep99} have 
$I_{\nu} \lesssim 4 \times 10^7$ Jy sr$^{-1}$. This limits exceeds the
measured far-IR background of 
$I_{\nu} ({\rm FIR}) \lesssim$ 1--2 $\times ~ 10^6$ Jy sr$^{-1}$ for
wavelengths longer than $\sim$ 10 $\mu$m \citep{fix98,hau98}.  
The known, finite sky brightnesses at optical and far-IR wavelengths
thus imply a turnover in the KBO number counts for
$R \gtrsim$ 30 \citep{kw01}.

Previous support for a turnover in the KBO number counts has relied on 
theoretical interpretations of available observations \citep[see][]{wl97}.
From numerical simulations, \cite{lev95} show that KBOs can excite an
eccentricity in the Pluto-Charon orbit.  If perturbations from KBOs are
the dominant source of the eccentricity, the measured $e$ yields an upper
limit to the number of KBOs with radii of 20--300 km.  Orbital integrations
of known Jupiter-family comets suggest an origin in the Kuiper Belt
\citep{dun88,lev94,dun95,dun97,ip97}.  
If the Kuiper Belt is the source of all Jupiter-family comets, the number
of known Jupiter-family comets and lifetimes derived from the orbital 
integrations provide limits on the number of KBOs with radii of 1--10 km.
These limits indicate that there are a factor of ten fewer KBOs with 
radii of 1--100 km than suggested by a simple extrapolation of 
equation (7) to $R \gtrsim$ 27.

The coagulation calculations provide more theoretical support for a
turnover in the number counts.  Models with fragmentation predict two
power-law size distributions, a merger population with $q$ = 3 at
large radii and a debris population with $q$ = 2.5 at small radii
\citep[Figures 6 and 9--10;][]{st97a,dav99,kl99a}.  The transition 
radius depends 
on the tensile strength of small objects.  For $S_0 \sim$ $10^3$ to 
$10^7$ erg g$^{-1}$, this radius is $\sim$ 1--100 km 
\citep{dav97,dav99,kl99a}, which agrees with the turnover 
radius derived from dynamical constraints.  

To place another constraint on the turnover radius, 
\citet{kw01} construct a physical model for the surface brightness
of small KBOs.  They adopt a broken-power law size distribution,
\begin{equation}
N_C (r) = \begin{array}{l l}
        n_0 (r/r_0)^{-q_1} & r > r_0 \\
                           & \\
        n_0 (r/r_0)^{-q_2} &  r \le r_0 \\
\end{array}
\end{equation}
\noindent
and assume objects lie in a ring around the Sun with surface density
$\Sigma \propto A^{-\gamma}$. The ring has an inner radius $A_1$ = 40 AU
and an outer radius $A_2$ = 50 AU.  The optical counts set $n_0$ and $q_1$.  
For an adopted $\omega_g$, $\mu_R$ results from a sum over all objects 
projected into a box with an area of 1 arcsec$^2$.  For thermal emission,
\citet{kw01} adopt the Backman \& Paresce (1993) relations to derive grain 
temperatures as a function of $A$ and sum the thermal emission from all
objects in a solid angle of 1 steradian.

\citet{kw01} demonstrate a clear turnover in the KBO number counts
(Figure 12).
Small KBOs with radii of 1 $\mu$m to $\sim$ 1 km must have a size 
distribution with $q \sim$ 3.4 or less to satisfy the known limits 
on the sky-surface brightness at optical and far-infrared wavelengths.

Figure 12 shows how the optical and 100 $\mu$m surface brightness increase 
with fainter KBO R-band magnitude.  Solid lines show results when all 
objects have $\omega_g$ = 0.04; dot-dashed lines show how the surface 
brightness changes when the albedo varies smoothly from $\omega_g$ = 0.04 
for $r \ge$ 1 km to $\omega_g$ = 0.5 for $r \le$ 0.1 km.  Larger albedos 
produce brighter optical surface brightnesses and a fainter far-IR surface 
brightness.  For models with $q_2$ = 3.5, KBOs with a small constant albedo 
have a limiting $\mu_R \sim$ 24.5 mag arcsec$^{-2}$, fainter than the 
observed sky brightness.  If small KBOs have $a_2$ = 3.5 and a large albedo, 
the predicted $\mu_R$ exceeds the observed background at $R \sim$ 70 mag. 
This limit corresponds to objects with $r \sim$ 0.03 mm.  In both cases,
the far-IR surface brightness exceeds the measured sky brightness
for $\lambda \le$ 240 $\mu$m at $R \approx$ 65--70 mag.  The predicted
far-IR surface brightness lies below measured limits at longer wavelengths.

A direct detection of diffuse light from KBOs would begin to provide
more stringent tests of coagulation models.  Measurements of the variation
of the diffuse light with ecliptic latitude or longitude would yield 
the scale height and orbital distribution of small KBOs.  The sensitivity
of archival deep HST WFPC2 images can improve constraints on the KBO 
optical background by a factor of ten. The {\it Space Infrared Telescope 
Facility} may improve the far-IR constraints by a similar factor. The
{\it Next Generation Space Telescope} will provide direct detections 
of individual KBOs near the proposed knee in the size distribution at 
$R \approx$ 28--31 mag and more accurate background measurements in 
the optical and near-IR.  These and other facilities will yield better
tests of model predictions for the size distribution of small KBOs.

\subsection{Radial distribution of KBOs}

The radial distribution of KBOs provides direct constraints on several 
physical processes in the outer solar system.  KBOs in the 2:1, 3:2,
and other orbital resonances yield information on dynamical interactions 
between small bodies and gas giant planets 
\citep[][Kuchner et al 2002]{hol93,dun95,hah99}.  KBOs in the 
scattered Kuiper Belt allow tests of models for the formation of 
the Oort comet cloud.  KBOs in the classical Kuiper Belt constrain the 
initial surface density and the formation history of large objects.  
Here, I concentrate on the radial distribution of classical KBOs, where 
coagulation models can offer some insight into the observations.

The observed radial distribution of KBOs in the classical Kuiper Belt
is uncertain.  Secular resonances with Neptune and Uranus truncate the inner
edge of the classical Kuiper Belt at $\sim$ 41 AU \citep{dun95}.  
Because the first surveys detected no KBOs outside 50 AU, \citet{don97} 
proposed an outer edge to the classical Kuiper Belt at $\sim$ 50 AU.  
Several large-angle surveys for KBOs provide support for an abrupt outer
edge at 48--50 AU \citep{jew98,all01,tru01}.  \citet{tr01} analyze 
discovery data for the apparent magnitude and heliocentric distance 
of all KBOs and derive an outer edge at 47 $\pm$ 1 AU.  They conclude
that plausible variations of the slope of the size distribution, the
maximum radius, and the albedo cannot produce the observed edge.
\citet{gla01} note that recent, unpublished surveys identify distant 
KBOs more frequently than older surveys, and conclude that the 
radial distribution of KBOs may continue smoothly beyond 48 AU.

Coagulation theory provides some explanations for possible origins of 
an outer edge to the observed radial distribution of classical KBOs.
Because the formation timescale for large objects depends on the 
orbital period, the size of the largest object is a sensitive function 
of heliocentric distance, $a$.  For $t <$ 100 Myr, multiannulus 
calculations yield $r_{max} \propto a^{-3}$ (Figure 7).  This result
implies a factor of two variation in the size of the largest object
from $a$ = 40 AU to $a$ = 50 AU.  After 100 Myr, the difference can
be (a) enhanced, if gravitational stirring by large objects at the
inner edge of the Kuiper Belt prevents the growth of objects farther
out in the Belt, or (b) diminished if gravitational stirring by
Neptune and other giant planets preferentially slows growth of
large objects at the inner edge of the Belt.  

The size distribution is an important factor in understanding the
reliability of an edge in the observed radial distribution of KBOs. 
Monte Carlo simulations of the observations demonstrate that the edge
is more robust for shallower input size distributions 
\citep{jew98,all01,tr01,gla01}.  For the
$q$ = 3 power law size distribution derived from single annulus
coagulation models, the edge is much more obvious than for the
$q$ = 3.25 power law derived from multiannulus calculations. 
Several test calculations suggest that $q$ grows with $a$. If
this conclusion holds with additional calculations, the coagulation
models favor steeper size distributions at larger distances in
the Kuiper Belt.  If this variation is real, the evidence for an
outer edge to the Kuiper Belt is more questionable.

Unless the tensile strength of objects decreases with $a$, the
variation of $r_{max}$ with $S_0$ from equation (5) is insufficient 
to yield a large variation in the radial distribution of KBOs.  
If $S_0$ is independent of $a$, $r_{max}$ changes by less than 
30\% at 40--50 AU.  \citet{tr01} show that this small 
change cannot produce the observed lack of KBOs beyond 48 AU.  
A factor of 10 change of $S_0$ at 40--50 AU can produce factor 
of 2--3 changes in $r_{max}$.  Because the magnitude of $S_0$
for KBOs is not well-known, quantifying changes of $S_0$ with
other variables in the model is pointless. Deriving tensile 
strengths of different comet families might help to quantify
possible variations of $S_0$ with $a$ (see below).

To make an initial theoretical prediction for the radial distribution
of classical KBOs based on the coagulation models, I use the 
number counts for multiannulus models from Figure 11 at 1 Gyr.  
The model assumes circular orbits, but
does not include collisional or dynamical evolution from 1 Gyr
to the present.  If this evolution is independent of $a$, then the 
model provides a reasonable first approximation to the present
situation in the outer solar system.  For simplicity, I quote the 
result of this model as a ratio, N(40--47 AU)/N(47--54 AU).  For
a limiting magnitude $R$ = 27, the multiannulus model with 
$S_0 = 2 \times 10^6$ erg g$^{-1}$ has N(40--47 AU)/N(47--54 AU) = 3; 
the model with $S_0 = 10^3$ erg g$^{-1}$ has N(40--47 AU)/N(47--54 AU) 
= 2.  Due to small number statistics, models with brighter limiting 
magnitudes produce unreliable results.  Because deeper surveys sample 
more of the size distribution, the number ratio declines as the
limiting magnitude increases.  

The results of the coagulation models suggest some caution in the
interpretation of the apparent edge in the radial distribution of
KBOs beyond 47 AU.  Factor of 2--3 declines in the apparent number 
of KBOs with $a$ are a natural outcome of coagulation models when
the input surface density follows a Minimum Mass Solar Nebula.
Larger changes are possible, if the surface density declines more
rapidly or with plausible changes to KBO properties as a function
of $a$.  Larger surveys to $R$ = 28 or deeper surveys to $R$ = 
29--30 should yield better statistics to discriminate among the 
possibilities.

A robust comparison between the models and observations is difficult 
due to uncertain observational biases and to uncertain long-term
dynamical evolution of the initial KBO population.  Most KBO surveys 
concentrate on regions near the ecliptic plane, where the success 
rate is larger; distant KBOs may have a different inclination 
distribution from nearby KBOs \cite{bro01}.  
The model estimates are smaller than the observed fraction,
N(40--47 AU)/N(47--54 AU) $\approx$ 4--6 \citep{tr01,gla01}.
The model assumes no migration 
in $a$ and no changes in $i$ from 1 Gyr to 5 Gyr; dynamical models
show that interactions with Neptune and other gas giant planets
change $a$, $e$, and $i$ on short timescales.  

\subsection{Orbital elements of KBOs}

The distributions of $e$ and $i$ yield information on the long-term 
dynamical evolution of KBOs.  Numerical integrations of KBO orbits
indicate that dynamical interactions with the gas giant planets 
dramatically change the orbital elements of objects in the outer
solar system
\citep[e.g.,][Kuchner et al. 2002]{tor90,hol93,dun95,mal96,lev97,mor97}. 
This gravitational sculpting of the KBO population produces several
dynamical KBO populations, including classical KBOs, plutinos
and other resonant KBOs, and scattered KBOs \citep[e.g.,][]{mal95,gla01}.
Understanding how these phenomena produce the current $e$ and $i$
distributions of KBOs remains a major puzzle.

Coagulation calculations provide an important foundation for
understanding the distributions of KBO orbital elements. Because
the giant planets are also condensing out of the solar nebula,
dynamical interactions between KBOs and gas giants are unimportant
during the early stages of KBO growth.  Collisional damping and
dynamical friction thus set the early velocity evolution of the 
KBO population.  These processes produce nearly circular  orbits
for large objects, $e \lesssim 0.001$ for $r_i \approx$ 100--1000 km,
and modestly eccentric orbits for smaller objects, $e \sim 0.01$ 
for $r_i \lesssim$ 10 km (Figures 9--10).  Once the collisional 
cascade begins, viscous stirring dominates the velocity evolution.
The orbits of all objects become more eccentric and more highly inclined.
After $\sim$ 1 Gyr, large objects in the multiannulus calculations 
have $e \sim$ 0.02 for $S_0$ = $10^3$ erg g$^{-1}$ models and 
$e \sim$ 0.1 for $S_0$ = $2 \times 10^6$ erg g$^{-1}$ models.
Both models have $i/e \approx$ 0.4.  These results indicate that
KBOs probably have significant $e$ and $i$ without dynamical
interactions with gas giant planets.

This conclusion is probably insensitive to initial conditions in
the Kuiper Belt.  \citet{kb01} show that 100--500 km objects can 
stir up velocities significantly on timescales of 1--5 Gyr in a 
Minimum Mass Solar Nebula.  Thus, large KBOs with sizes of 
500--1000 km can stir up other KBOs to large $e$ and $i$ on a
1 Gyr timescale.  If KBOs form in a low mass solar nebula,
stirring timescales are longer,
$\sim$ 5 Gyr for 1000 km objects with 10\% of the Minimum Mass
and $\sim$ 50 Gyr for 1000 km objects with 1\% of the Minimum Mass.
Thus, our scenario for producing KBOs in a Minimum Mass Solar Nebula
leads to KBOs with large $e$ and $i$.  Models which form massive KBOs 
in a low mass solar nebula yield KBOs with low $e$ and $i$.

These results indicate that gravitational sculpting and the internal
dynamics of KBOs are important in creating the current distributions
of $a$, $e$, $i$ for KBOs\footnote{Gravitational interactions with
passing stars can also modify the orbital elements of KBOs \citep{ida00b}.}. 
Viscous stirring between large KBOs broadens
the $e$ and $i$ distributions with time; gravitational sculpting by
the gas giants broadens the $e$ and $i$ distributions {\it and} selects 
stable ranges of $e$ and $i$.  Careful treatment of both processes is 
necessary to understand the current orbital elements of KBO populations.

\section{DISCUSSION AND SUMMARY}

The discovery of the Kuiper Belt in the 1990's provides fundamental
constraints 
on models for the formation and evolution of planets in the outer parts
of our solar system.  The observations imply $\sim 10^5$ KBOs with radii 
of 50--500 km and a total mass of $\sim$ 0.1--0.2 $M_{\oplus}$ beyond
the orbit of Neptune.  The theoretical challenge is to understand the
formation of large objects in a current reservoir of material that is
$\sim$ 1\% of the initial mass in the solar nebula.  This goal assumes 
that KBOs formed locally and that the initial surface density of the 
solar nebula did not decease abruptly beyond the orbit of Neptune.
Observations indicate typical disk radii of at least 100--200 AU in 
nearby pre-main sequence stars, which suggests that the disk of our 
solar system originally continued smoothly beyond the orbit of Neptune.   
Testing the assumption of local KBO formation relies on future comparisons
between observations and theory.

Coagulation calculations appear to meet the challenge posed by KBOs.
Published numerical calculations demonstrate that the formation of 
a few Plutos and numerous 100--500 km KBOs in the outer parts of a 
solar system is inevitable \citep[Stern 1995, 1996a;][]{st97a,dav99,kl99a}.  
For a variety of initial conditions, collisions between small bodies 
at 30--50 AU naturally produce larger objects.  Once there is a range 
in sizes, dynamical friction efficiently reduces the orbital 
eccentricities of the largest objects.  Large objects in nearly 
circular orbits grow quickly.  At 30--50 AU, runaway growth can 
produce 100 km and larger objects on short timescales.  These objects 
then grow slowly to radii of 1000 km or more.  

The initial disk mass sets the timescale for Pluto formation in the
outer parts of a solar system.  Objects grow faster in more massive disks.  
For single annulus calculations of planetesimals orbiting the Sun,
the timescale to produce the first Pluto is
\begin{equation}
t_P \approx 20 ~ {\rm Myr} ~ \left ( \frac{\Sigma_{35}}{0.2 ~ {\rm g ~ cm^{-2}}} \right )^{-1} ~ ,
\end{equation}
where $\Sigma_{35} \approx 0.2 ~ {\rm g ~ cm^{-2}}$ is the initial surface 
density of a minimum mass solar nebula model extrapolated into the 
Kuiper Belt at $\sim$ 35 AU \citep[Figure 2; see also][]{st97a,kl99a}.  
This timescale depends weakly on the initial conditions.  Growth is
more rapid in a solar nebula with small initial eccentricities and
with small initial bodies \citep{kl99a}.  

The growth timescale in the Kuiper Belt is smaller than expected from
coagulation calculations in the inner solar system.  \citet{lis96} 
estimate a timescale to produce Moon-sized ($10^{26}$ g) objects as

\begin{equation}
t_M \approx {\rm 0.5 ~ Myr} \left ( \frac{{\rm 1~g~cm^{-2}}}{\Sigma(a)} \right ) \left ( \frac{a}{{\rm 1 ~ AU}} \right ) ^{3/2} ~ .
\end{equation}
\noindent
This relation implies timescales of $\sim$ 500 Myr at 35 AU and
$\sim$ 1 Gyr at 45 AU.  Our single annulus models yield 
$t_M \sim$ 100 Myr at 35 AU and
$t_M \sim$ 600 Myr at 70 AU. 
For calculations where the initial size distribution is composed of
1--10 km bodies, multiannulus models imply $t_M \sim$ 200--300 Myr at 
40--50 AU.  Collisional damping causes the difference between our 
results and equation (13). In our calculations, collisional damping 
between small objects with radii of 1 m to 1 km reduces eccentricities 
by factors of 5--10.  Dynamical friction couples the eccentricity 
reduction of the small bodies to the largest bodies. Because runaway
growth begins when gravitational focusing factors are large,
collisional damping in our Kuiper Belt models leads to an early 
onset of runaway growth relative to models of the inner solar system
where the collisional evolution of small bodies is not important.  

Once large objects form in the outer part of a solar system, they
stir up the velocities of small objects with radii of 10--100 km
or less.  Velocity stirring retards growth and produces debris.
When the collision energy of small bodies is comparable to their 
tensile strength, the small bodies undergo a collisional cascade
where planetesimals are ground down into smaller and smaller objects.
This process produces numerous small grains which are ejected by
radiation pressure ($\lesssim$ 1--3 $\mu$m grains) or pulled towards 
the Sun by Poynting-Robertson drag ($\gtrsim$ 1--3 $\mu$m grains).  
These grains are 
lost on short timescales of 1 Myr or less.  When the collisional 
cascade begins, most of the mass in the outer solar system is 
contained in small objects that are easy to fragment.  The collisional 
cascade thus robs the larger bodies of material.  Because collisional 
cascades start sooner in the evolution when bodies are weaker, 
the size of the largest object in a calculation depends on the 
tensile strength of the small planetesimals.  Our models yield
Earth-sized objects in the Kuiper Belt for $S_0$ = $2 \times 10^6$
erg g$^{-1}$ and Pluto-sized objects for $S_0$ = $10^2$ to $10^3$ erg g$^{-1}$.

The theoretical models thus resolve the dilemma of large objects in a 
low mass Kuiper Belt.  Runaway growth of small objects at 40--50 AU in 
the solar nebula places $\sim$ 5\%--10\% of the initial mass in large
objects with radii of 50--500 km or larger.  The collisional cascade converts
80\%--90\% of the initial mass into debris which is removed from the
Kuiper Belt on short timescales.  Over the 4.5 Gyr lifetime of the
solar system, gravitational interactions between KBOs and the gas giant
planets can remove $\sim$ 50\% to 80\% of the remaining mass.  Given 
the uncertainties, collisions and dynamics appear capable of removing
more than 90\% of the original mass in the Kuiper Belt.

The observed size distribution of KBOs provides strong observational 
tests of this picture.  The final size distribution of a Kuiper Belt 
calculation has three components.  The merger component at large sizes 
is a power-law with $q_f \approx$ 3.0--3.5; the debris component at small 
sizes is a power law with $q_f \simless$ 2.5.  The collisional cascade 
depletes objects with intermediate sizes of 0.1--10 km.  Depletion 
produces a dip in the size distribution for $S_0 \lesssim 10^5$ erg g$^{-1}$.

The observations of large KBOs generally agree with the power law 
slope predicted for the merger component.  The data are consistent 
with $q$ = 3.3--3.5; the multiannulus models predict $q$ = 3.15--3.35.  
If dynamical interactions and collisional evolution continue 
to remove KBOs from the 40--50 AU annulus after 1 Gyr, 
the predicted number of KBOs is within a factor 
of two of the observed number of KBOs.  The multiannulus calculations 
produce more KBOs with radii of 1000 km or larger than are observed 
with current surveys.  The predicted number of these 
large objects depends on $S_0$ and is therefore uncertain.  The 
observed number of large objects is plagued by small number statistics.
Future surveys will provide robust constraints on the population of
large objects.  Improved multiannulus coagulation calculations which 
include dynamical interactions with gas giant planets will improve the
predictions.

Current constraints on the population of small KBOs are also consistent
with model predictions.  The data indicate a turnover in the KBO number
counts, which implies a turnover in the size distribution for small
objects.  The derived turnover radius of 0.1--10 km
is close to theoretical predictions.
Better observations of the optical and far-IR surface brightnesses of 
the Kuiper Belt can provide better estimates of the slope of the size 
distribution for KBOs with radii of 1 mm to 1 m.  Observations with 
larger telescopes may detect the turnover radius directly.

Measuring the tensile strengths of comets provides an interesting test 
of this picture of KBO formation.  In our models, the formation of Pluto 
by coagulation requires a tensile strength $S_0 \gtrsim$ 400 erg g$^{-1}$.  
Large tensile strengths, $S_0 \gtrsim$ $10^5$ erg g$^{-1}$, allow the
formation of large bodies, $\sim$ 2000--3000 km, which have not been 
detected in the outer solar system.  Because objects with radii of
2000--3000 km can form before Neptune reaches its current mass, the 
lack of large KBOs implies $S_0 \lesssim$ $10^4$ erg g$^{-1}$ in the 
coagulation theory.  Estimates on the tensile strength derived from
comet Shoemaker-Levy 9, $S_0 \sim$ $10^2$ erg g$^{-1}$ to $10^4$ erg g$^{-1}$
\citep[e.g.][]{gre95}, are close to the lower limit required to form Pluto.
Theoretical estimates have a much larger range, $S_0 \sim$ $10^2$ 
erg g$^{-1}$ to $10^6$ erg g$^{-1}$ \citep{sir00}.  As theoretical
estimates improve and observations of disrupted comets become more 
numerous, these results can constrain the coagulation models.

The coagulation calculations demonstrate that planet formation in the 
outer parts of other solar systems is also inevitable.  The mass of a 
Minimum Mass Solar Nebula is comparable to the median disk mass derived 
for nearby pre-main sequence stars \citep{bec99,lad99,man00}.  The
formation timescale for a 1000 km planet at 30--50 AU in one of these
disks is therefore $\sim$ 10--30 Myr.  Although this planet cannot be 
observed directly, gravitational stirring leads to a collisional
cascade and copious dust production.  In the multiannulus models,
dust is produced at a rate of roughly 0.1--1 Earth mass every 100 Myr
\citep[see also][]{kb02}. 

Observations of nearby debris disk systems are consistent with dust
produced in a planet-forming disk.  The sizes of debris disks,
$\sim$ 10--1000 AU, are similar to the radius of the Kuiper Belt.
The ages of the youngest debris disk systems are comparable to the 
Pluto formation timescale of $\sim$ 10--20 Myr \citep{lag00}.  If 
the timescale for Poynting-Robertson drag sets the residence time 
for 1 $\mu$m and larger dust grains in the disk, the instantaneous 
dust mass in the disk is $\sim$ 0.1--1 lunar masses.  This mass is 
comparable to the dust masses inferred from IR observations of debris 
disk systems such as 
$\alpha$ Lyr and $\beta$ Pic \citep{bac93,lag00}.  Finally, the
duration of the collisional cascade in our Kuiper Belt models,
$\sim$ 100 Myr to $\sim$ 1 Gyr, is similar to the estimated lifetimes
of debris disk systems, $\sim$ 500 Myr \citep[][2001]{hab99}.
\citet{kb01} derive a similar predicted lifetime for debris disk
systems from the coagulation equation \citep[see also][]{ken00}.

To make the connection between KBOs and debris disks more clear,
\citet{ken99} investigate planet formation in the dusty ring of
HR 4796A \citep{jay98,koe98,au99,sch99,gr00}.  They show that 
a planetesimal disk with a mass of
10--20 times the mass of the Minimum Mass Solar Nebula can form
a dusty ring on 10--20 Myr timescales, comparable to the estimated
age of HR 4796A.  The model ring has a radial optical depth $\sim$ 1,
in agreement with limits derived from infrared images and from the 
excess infrared luminosity.  Although the initial mass in this 
single annulus calculation is large, multiannulus calculations 
suggest similar timescales with much smaller masses.

Finally, multiannulus calculations are an important new tool in
developing a robust model for planet formation.  Current computer 
technology allows practical multiannulus calculations that cover 
roughly a decade in disk radius. We are thus 1--2 orders of 
magnitude from constructing model grids of complete solar systems.
Faster computers should resolve this difficulty in the next few years
and allow us to consider the interfaces between (i) gas giants and 
terrestrial planets and (ii) gas giants and the Kuiper Belt.  With 
some limitations, current multiannulus calculations promise 
predictions for the radial variation of the disk scale height \citep{kb01} 
and the disk luminosity \citep{kb02} as a function of stellar age, 
disk mass, and other physical parameters.  Detailed comparisons 
between these predictions and observations of debris disks will 
yield interesting constraints on the physics of planet formation
in other solar systems.  Applying these results to our solar system
will provide a better idea how the Earth and other planets in our
solar system came to be.

\vskip 4ex

I thank J. Luu for suggesting our joint projects and B. Bromley 
for advice and assistance in preparing the coagulation code for 
a modern, parallel computer.
The JPL and Caltech supercomputer centers provided generous
allotments of computer time through funding from the NASA Offices 
of Mission to Planet Earth, Aeronautics, and Space Science.  Advice 
and comments from M. Geller, M. Kuchner, C. Lada, B. Marsden, 
R. Windhorst, and J. Wood greatly improved the content and the 
presentation of this review.

\vfill
\eject

\clearpage


\epsfxsize=8.0in
\hskip -10ex
\epsffile{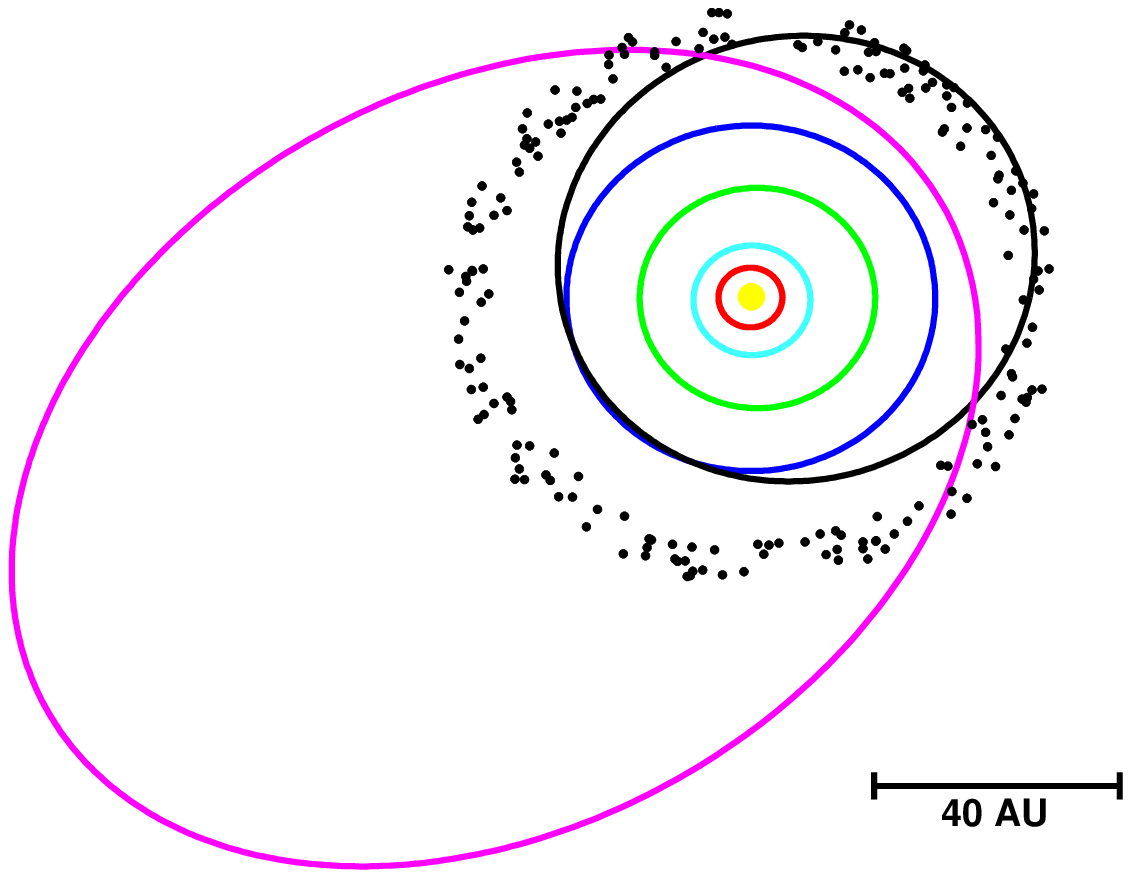}
\figcaption[fg1.ps]
{Top view of the solar system.  The yellow filled circle is
the Sun.  Colored ellipses indicate the orbits of Jupiter 
(dark purple), Saturn (cyan), Uranus (green), Neptune (blue),
Pluto (black), and the scattered Kuiper Belt object 1996 TL$_{66}$
(magenta).  The black dots represent 200 classical Kuiper Belt 
objects randomly distributed in a band between 42 AU and 50 AU.
The bar at the lower right indicates a distance of 40 AU.}

\epsfxsize=7.0in
\epsffile{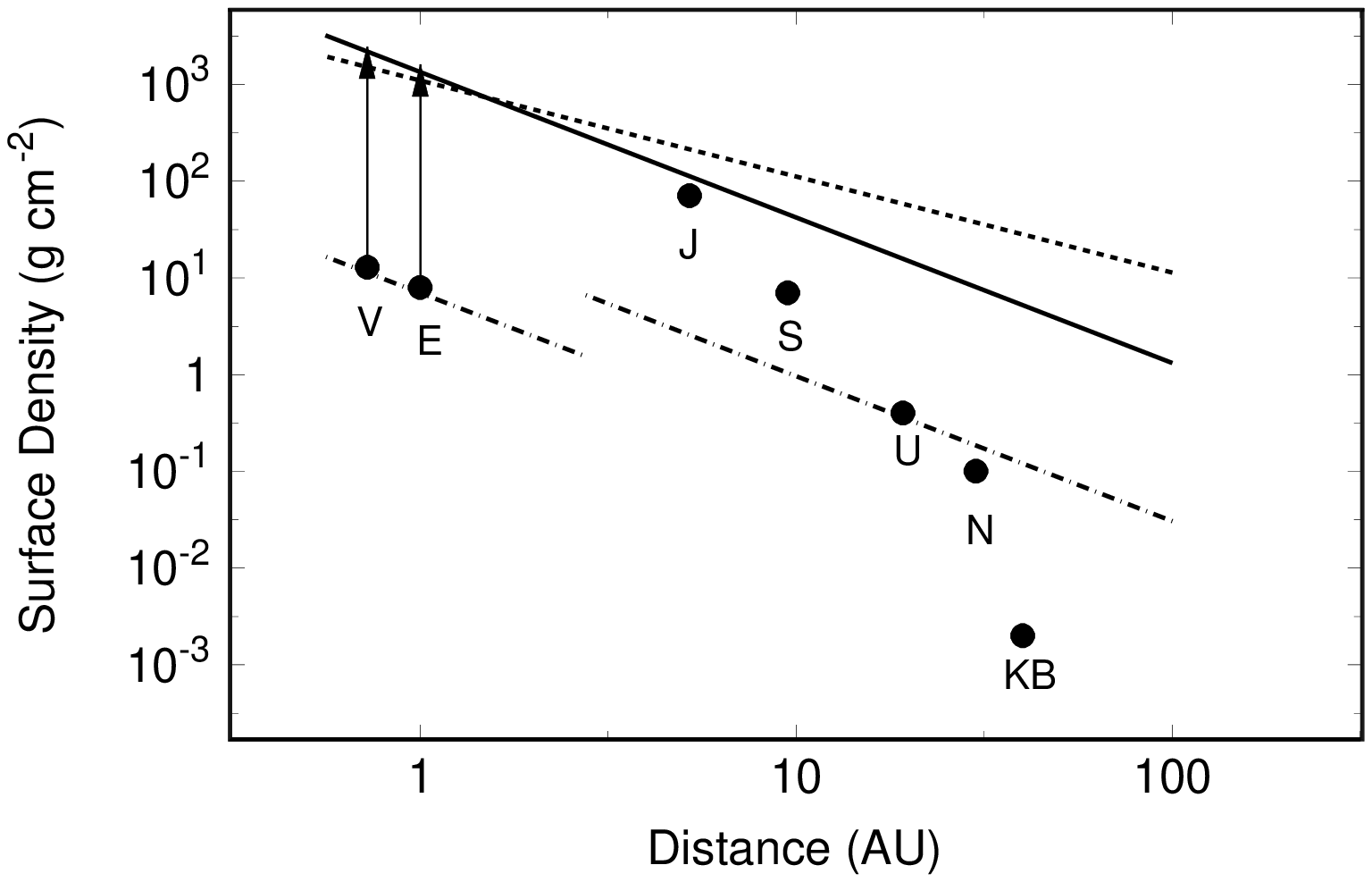}
\figcaption[fg2.eps]
{Surface density distribution in the solar system,
assuming that the mass is spread uniformly over an annulus centered on 
the orbit of the planet.  The arrows indicate the surface density
for terrestrial planets if augmented to a solar abundance of hydrogen 
and helium. The solid and dot-dashed curves indicate $\Sigma \propto A^{-3/2}$
\citep{wei77a,hay81}; the dashed line indicates
$\Sigma \propto A^{-1}$ \citep{cam95}.}

\epsfxsize=7.0in
\hskip -10ex
\epsffile{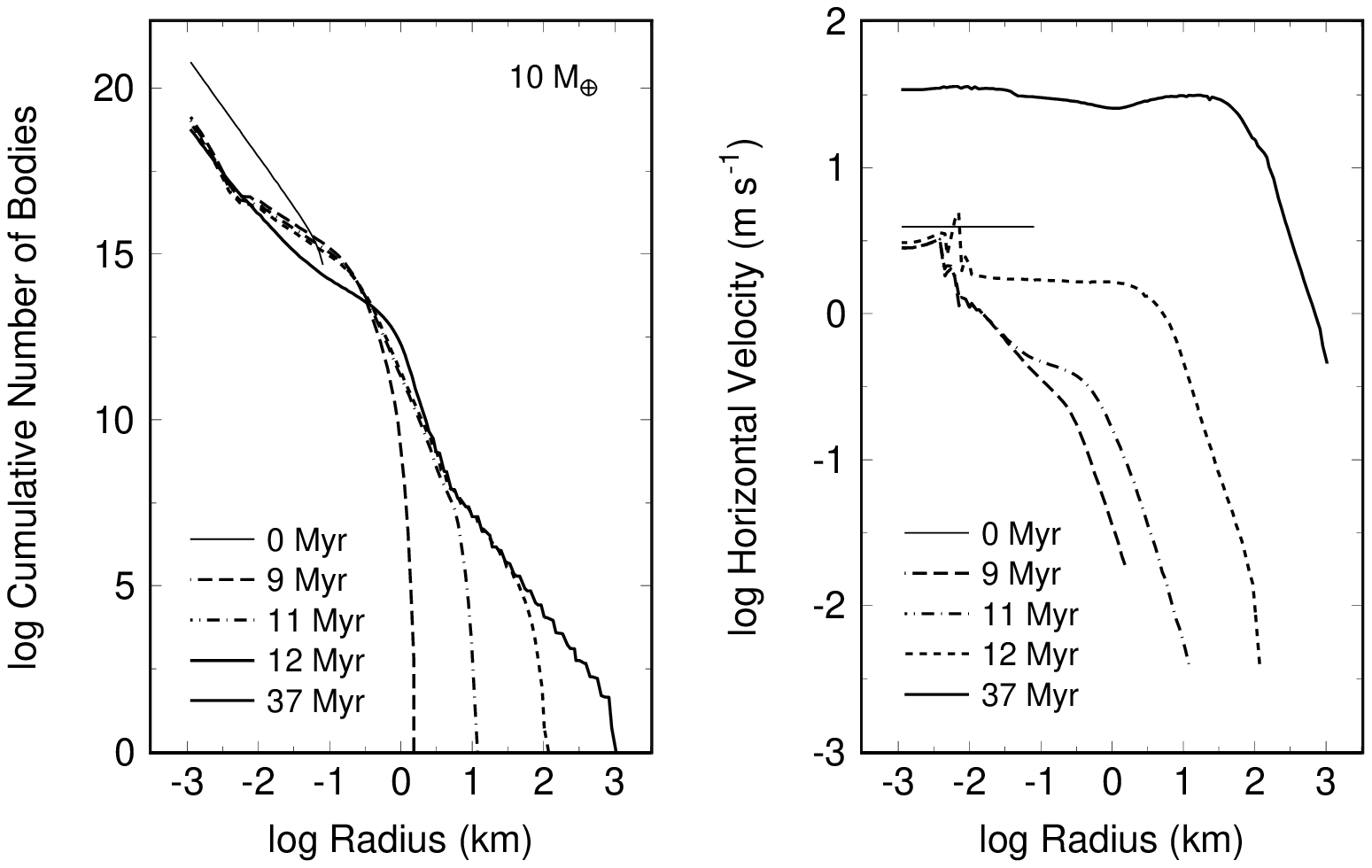}
\figcaption[fg3.eps]
{Evolution of a single annulus coagulation model with $M_0$ = 10 $M_{\oplus}$, 
$e_0 = 10^{-3}$, and $S_0 = 2 \times 10^6$ erg g$^{-1}$: 
(a) cumulative size distribution (left panel), and (b)
horizontal velocity (right panel) as a function of time. 
Collisional growth is
quasi-linear until the largest bodies have $r_{max}$ = 1--2 km
at 9--10 Myr. Collisional damping reduces the velocities of all
bodies to $\sim$ 1--2 m s$^{-1}$ on this timescale; dynamical 
friction damps the velocities of larger bodies to $\sim 10^{-2}$
m s$^{-1}$.  Runaway growth then produces objects with radii
of 100 km in another 2--3 Myr.  Viscous stirring heats up 
particle velocities as objects grow to sizes of 100--300 km.
Runaway growth ends.  A prolonged oligarchic growth phase leads 
to the production of 1000 km objects; the horizontal velocities
are then $\sim$ 30--40 m s$^{-1}$ for the smallest objects and
$\sim$ 1 m s$^{-1}$ for the largest objects. Adapted from \citet{kl99a}}

\epsfxsize=8.0in
\epsffile{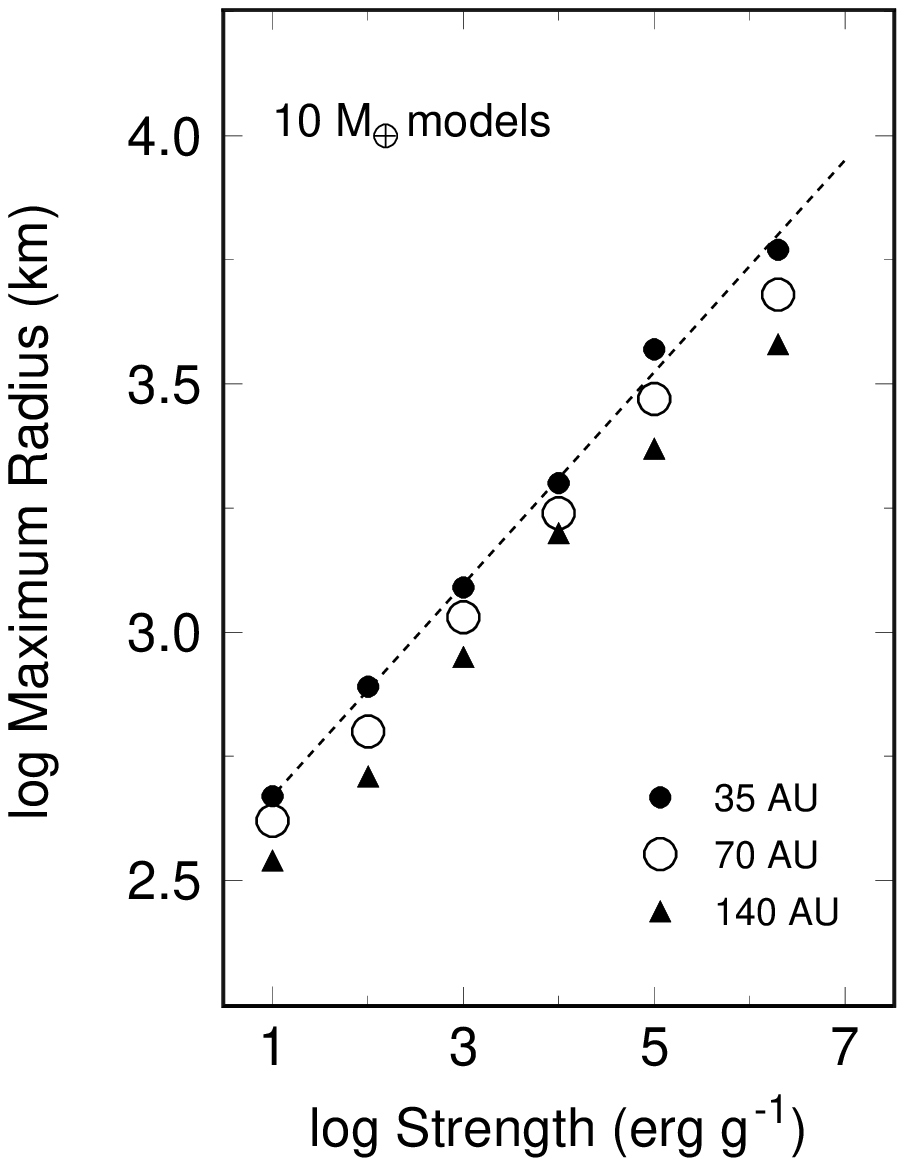}
\figcaption[fg4.eps]
{Maximum radius as a function of tensile strength and heliocentric distance
for single annulus models with 
$M_0 = 10~M_{\oplus}$ and $e_0 = 10^{-4}$.  At a given
heliocentric distance,  larger planets grow from stronger planetesimals.
At a given tensile strength, smaller planets form at larger heliocentric
distances. Adapted from \citet{kl99a}.}

\epsfxsize=8.0in
\hskip -10ex
\epsffile{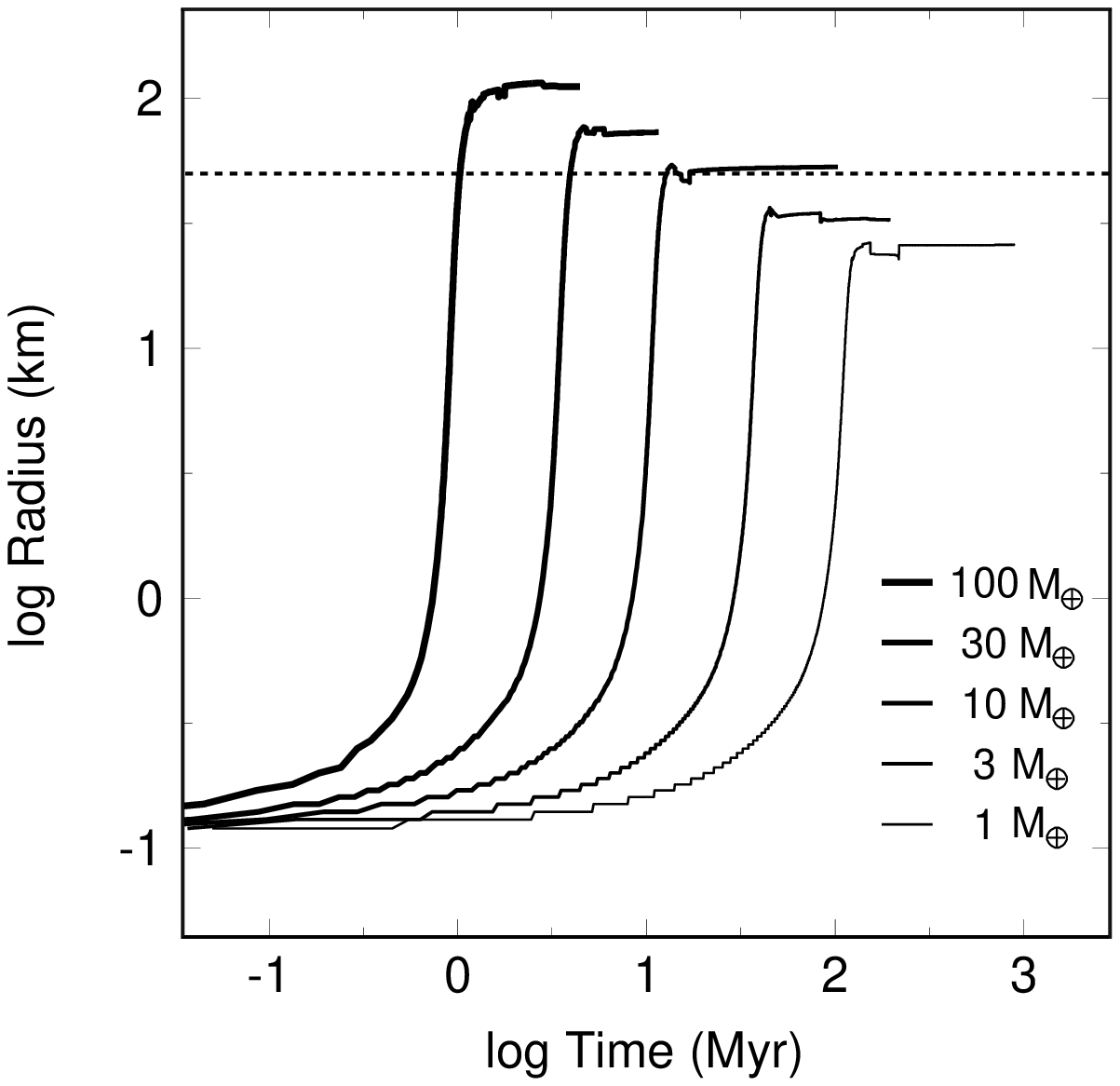}
\figcaption[fg5.eps]
{Evolution of $r_5$, the radius where the cumulative number of objects
is $10^5$, with time as a function of initial mass, $M_0$, for 
single annulus models with $e_0 = 10^{-3}$.  The horizontal dashed line
indicates the constraint on $r_5$ set by current observations. 
Adapted from \citet{kl99a}.}

\epsfxsize=6.50in
\hskip -7ex
\epsffile{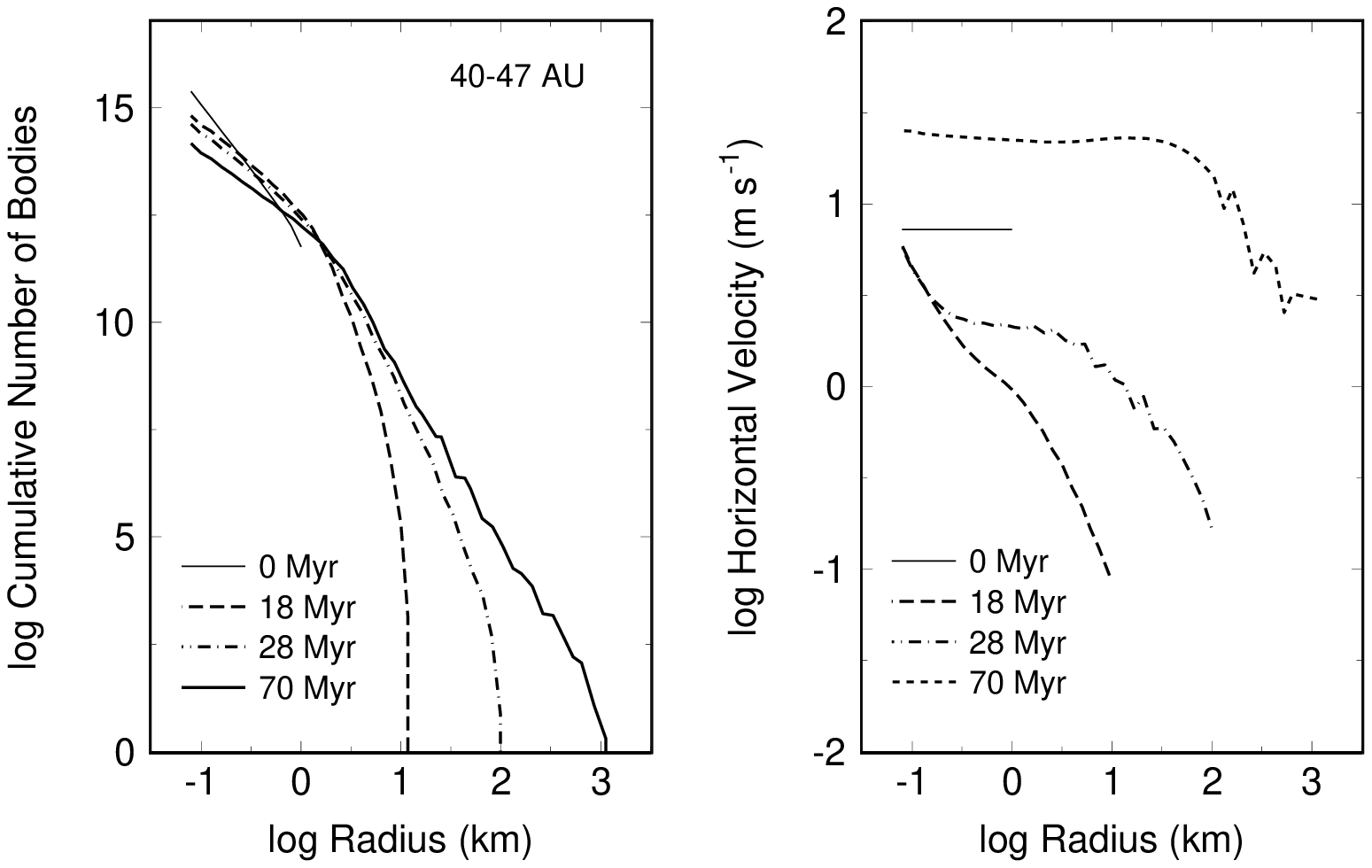}
\vskip -2ex
\figcaption[fg6.ps]
{Evolution of a multiannulus coagulation model with 
$\Sigma_i$ = 0.3 ($a_i$/35 AU)$^{-3/2}$, $e_0 = 2 \times 10^{-3}$, 
$S_0 = 2 \times 10^6$ erg g$^{-1}$, and velocity evolution: 
(a) cumulative size distribution (left panel), and (b) horizontal 
velocity (right panel)as a function of time. Collisional growth is
quasi-linear until the largest bodies have $r_{max}$ = 3--10 km
at 20 Myr. Collisional damping reduces the velocities of small
bodies to $\sim$ 1--5 m s$^{-1}$ on this timescale; dynamical 
friction reduces the velocities of larger bodes to $\lesssim 10^{-1}$
m s$^{-1}$.  Runaway growth then produces objects with radii of 
100 km in 10 Myr.  Viscous stirring increases particle velocities 
as objects grow to sizes of 300--500 km, and runaway growth ends.  
An oligarchic growth phase leads to the production of 1000 km 
objects after $\sim$ 70 Myr; the horizontal velocities are then $\sim$ 
40--50 m s$^{-1}$ for the smallest objects and $\sim$ 2--3 m s$^{-1}$ 
for the largest objects.}

\epsfxsize=8.0in
\hskip -5ex
\epsffile{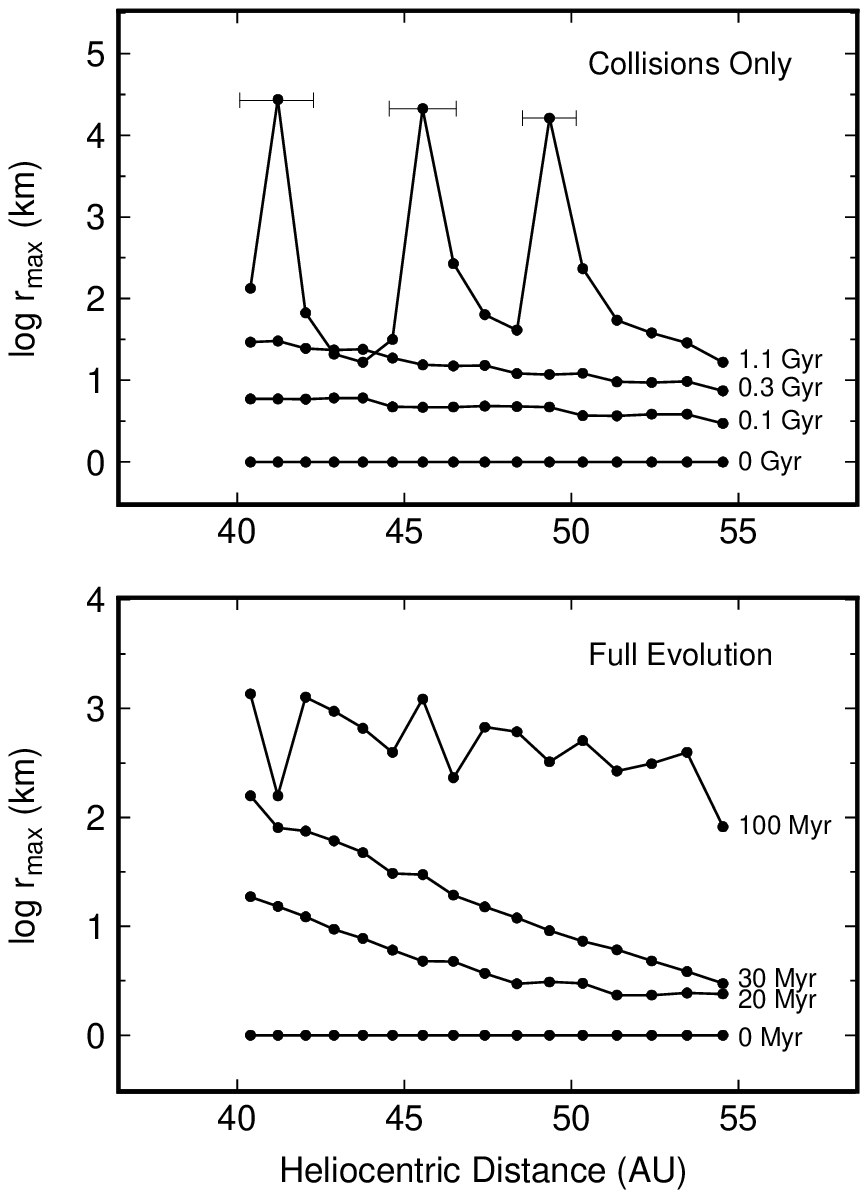}
\figcaption[fg7.eps]
{Mass of the largest body in each annulus of two multiannulus coagulation
calculations.
(a) top panel: the radius of the largest object at 0, 0.1, 0.3, and 1.1 Gyr
for a model without velocity evolution;
(b) bottom panel: the radius of the largest object at 0, 20, 30, and 100 Myr
for a model with velocity evolution.  Each annulus initially contains 
0.1--1 km bodies with the surface density in solid material equivalent to
a Minimum Mass Solar Nebula.  Objects grow faster in models with velocity
evolution, but objects become larger in models without velocity evolution.
The error bars in the top panel indicate the Hills radius $R_H$ for each 
large body formed in the calculation without velocity evolution.  
Objects cannot accrete material beyond 2.4 $R_H$ 
\citep[see also][]{ale98,kok98}.}

\epsfxsize=8.0in
\hskip -7ex
\epsffile{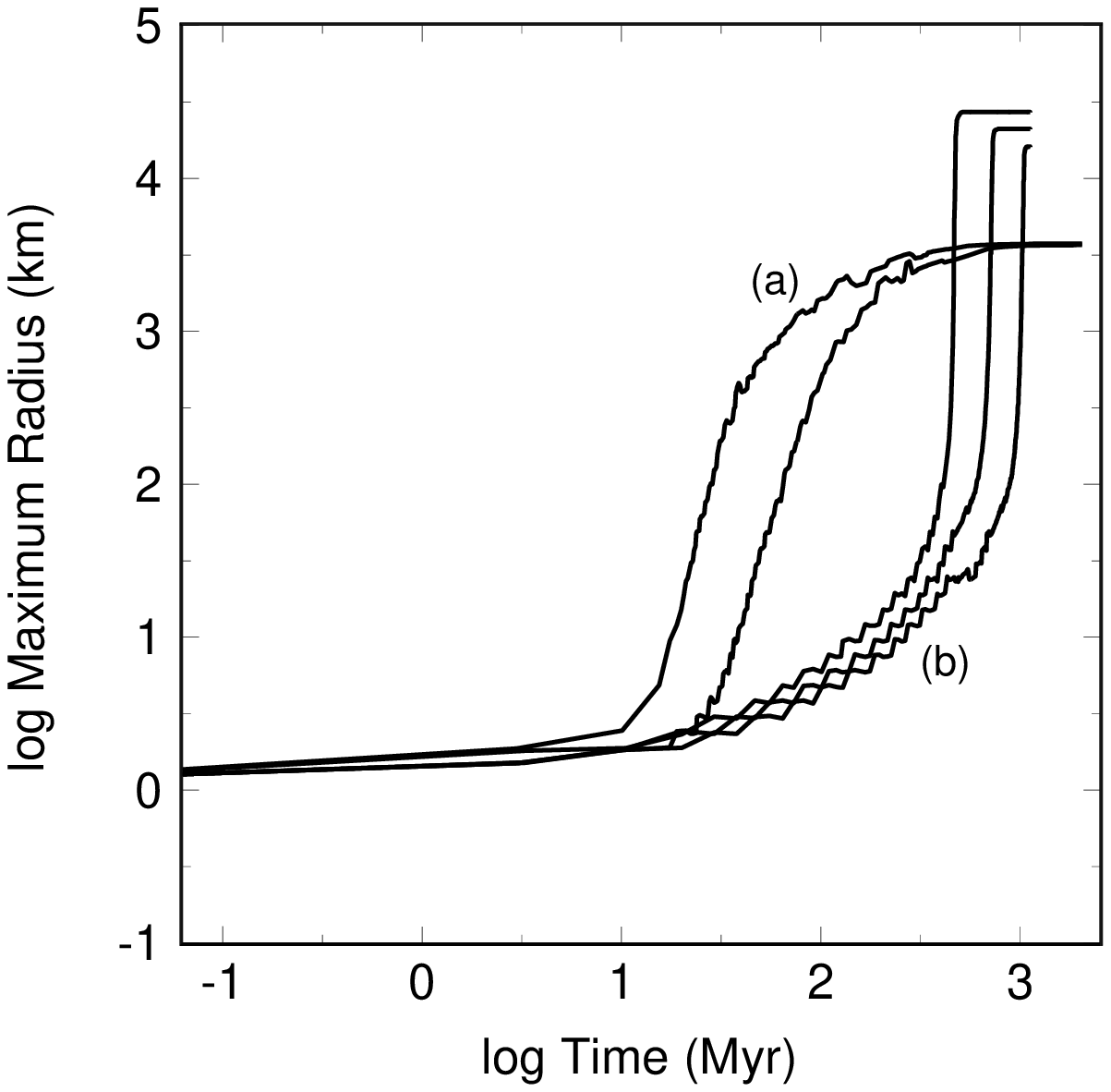}
\vskip -3ex
\figcaption[fg8.eps]
{Evolution of the largest objects in several multiannulus calculations
of planetesimal evolution at 40--55 AU.  The two curves labeled `(a)' show
the growth of the largest objects in a multiannulus calculation with
velocity evolution and fragmentation (Figure 6).  The first curve
plots the growth of the largest object in annulus (1); the second 
curve plots the growth of the largest object in annulus (15).  The
curves labeled `(b)' show the growth of the largest objects in a
calculation with fragmentation but no velocity evolution (Figure 7).
The largest object in annulus (2) reaches runaway growth before 
the largest object in annulus (7), which achieves its maximum radius
before the largest object in annulus (11) begins runaway growth.}

\epsfxsize=7.0in
\hskip -10ex
\epsffile{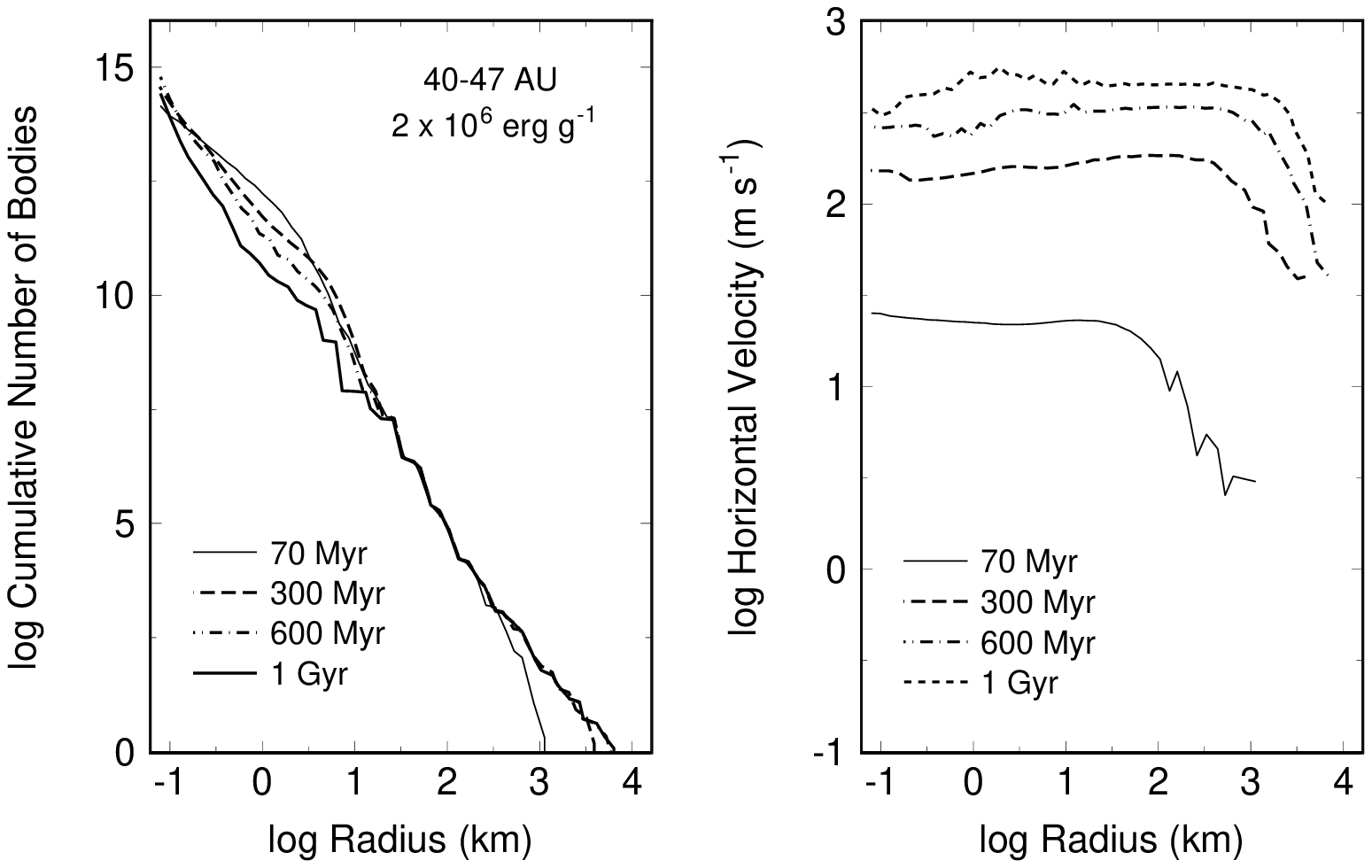}
\figcaption[fg9.eps]
{Late evolution of a multiannulus model with $\Sigma_i$ = 0.3 
($a_i$/35 AU)$^{-3/2}$, $e_0 = 2 \times 10^{-3}$, 
$S_0 = 2 \times 10^6$ erg g$^{-1}$, and velocity evolution: 
(a) cumulative size distribution (left panel), and (b)
horizontal velocity (right panel) as a function of time.  After the first
Pluto-sized object forms at $\sim$ 70 Myr, growth is oligarchic. 
As objects grow from radii of $\sim$ 1000 km to radii of $\sim$ 6000 km,
viscous stirring increases particle velocities to the shattering limit.
Shattering reduces the population of 1--10 km objects on timescales
of 500 Myr to 1 Gyr.}

\epsfxsize=7.0in
\hskip -10ex
\epsffile{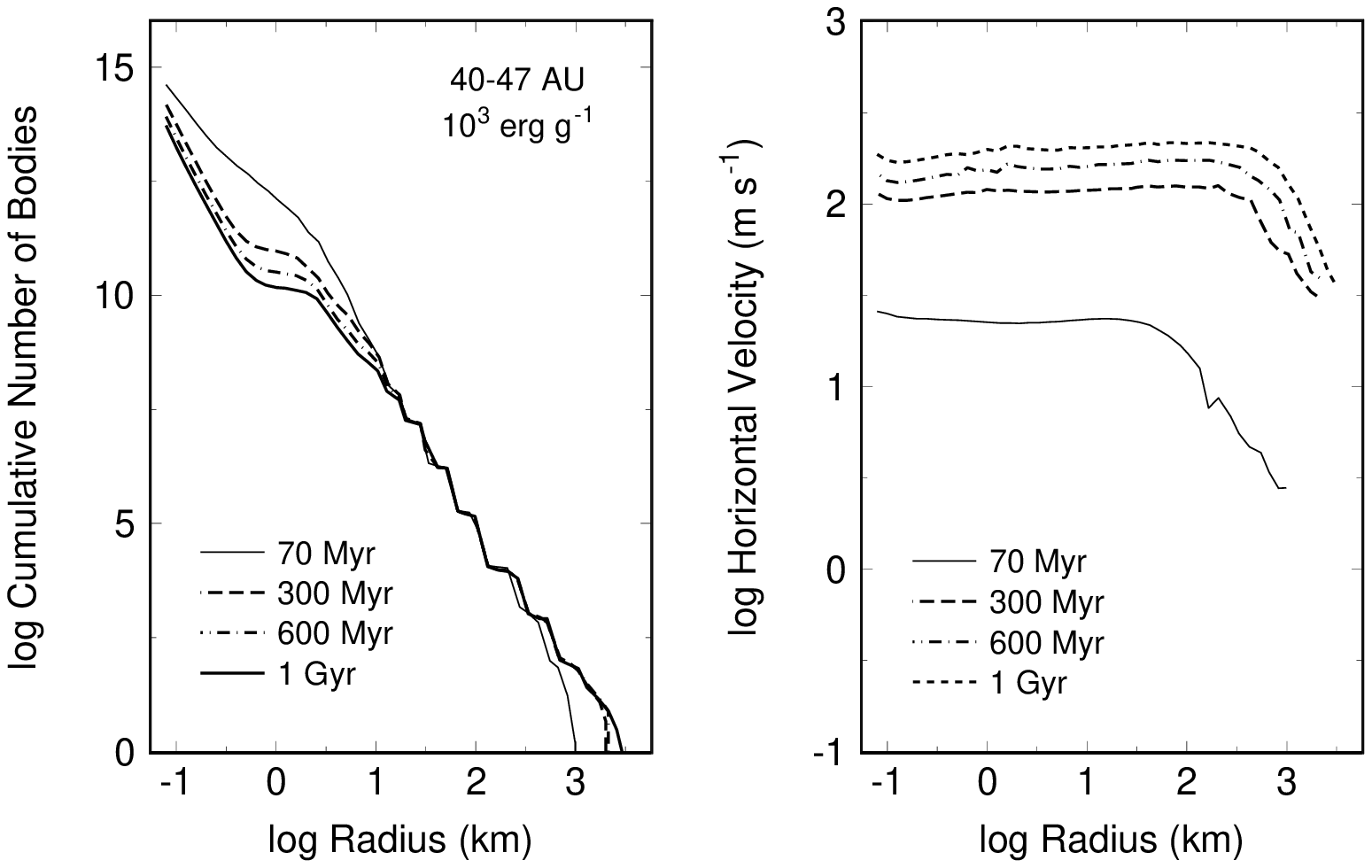}
\figcaption[fg10.eps]
{Late evolution of a multiannulus model with $\Sigma_i$ = 0.3 
($a_i$/35 AU)$^{-3/2}$, $e_0 = 2 \times 10^{-3}$, 
$S_0 = 10^3$ erg g$^{-1}$, and velocity evolution: 
(a) cumulative size distribution (left panel), and (b)
horizontal velocity (right panel) as a function of time.  After the first
Pluto-sized object forms at $\sim$ 70 Myr, growth is oligarchic. 
As objects grow from radii of $\sim$ 1000 km to radii of $\sim$ 3000 km,
viscous stirring increases particle velocities to the shattering limit.
At times of 300 Myr to 1 Gyr, shattering reduces the population of 
small objects and produces a prominent dip in the size distribution 
at 0.3--3 km.}

\epsfxsize=7.7in
\epsffile{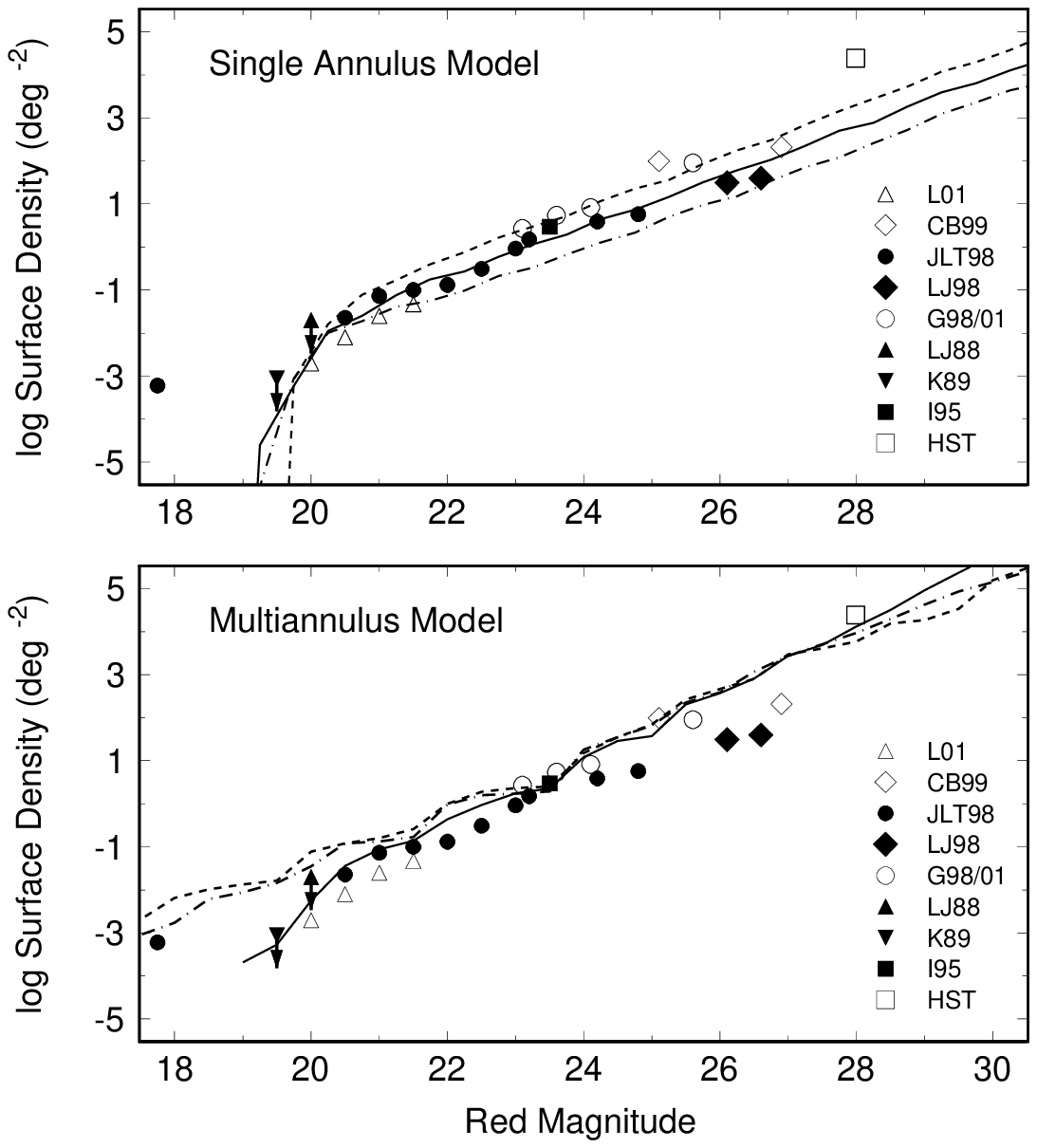}
\vskip -2ex
\figcaption[fg11.ps]
{Comparison of model luminosity functions of KBOs with observations.  
Data are as indicated in the legend of each panel and are from
Cochran et al. (1998; HST), Irwin et al. (1995; I95), 
Kowal 1989 (1989; K89), Luu \& Jewitt (1988; LJ88),
Gladman et al. (1998, 2001; G98/01), Luu \& Jewitt (1998; LJ98), 
Jewitt et al. (1998; JLT98), Chiang \& Brown (1999; CB99), and
Larsen et al. (2001; L01).  
Error bars for each datum -- typically a factor of 2--3 -- and 
the upper limit from Levison \& Duncan (1990) are not shown 
for clarity. The lines plot luminosity functions for 
(a) upper panel: single annulus models at 35 AU with $e_0 = 10^{-3}$ and 
$M_0 \approx$ 0.3 (dot-dashed), 1.0 (solid), and 3.0 (dashed) times the
Minimum Mass Solar Nebula and 
(b) lower panel: multiannulus models at 40--55 AU for models with
a mass in solids of a Minimum Mass Solar Nebula with 
$e_0 = 2 \times 10^{-3}$ and $S_0$ = $2 \times 10^6$ erg g$^{-1}$
(solid line: 70 Myr; dashed line: 1 Gyr) and $S_0$ = $10^3$ erg g$^{-1}$ 
(dot-dashed line: 1 Gyr).}

\hskip 10ex
\epsfxsize=8.0in
\epsffile{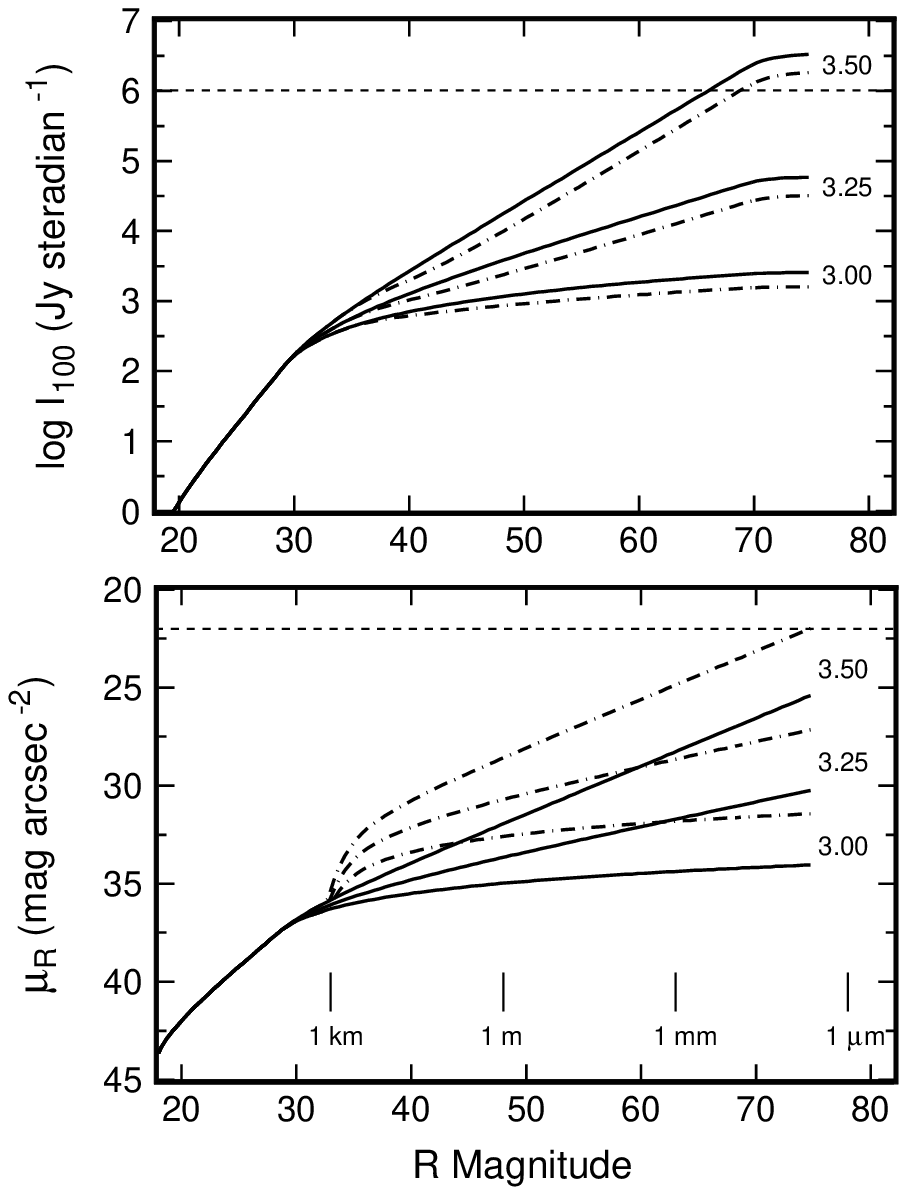}
\figcaption[fg12.eps]{ Far-infrared and optical surface 
brightness as a function of R magnitude for a physical model of
KBO grains.  The model assumes a broken power law size distribution,
equation (11), albedo $\omega_g$, and a surface density distribution 
for KBOs in a ring at 40--50 AU.  Solid curves show results for 
$a_1$ = 3, $\omega$ = 0.04, and $a_2$ as indicated at the right 
end of each curve.  Dot-dashed curves repeat this model for small
grains with larger $\omega$. Each model is consistent with 
observations of the optical counts at $R \le$ 26--27.}

\end{document}